\documentclass[aps,prd,amssymb,eqsecnum,showpacs] {revtex4}
\usepackage{graphicx}
\usepackage{psfrag}

\begin{document}

\title{Strategies for the Characteristic Extraction of Gravitational Waveforms}

\author{M.~C. Babiuc${}^{1}$, N.~T. Bishop${}^{2}$, B. Szil\'{a}gyi${}^{3,4}$
and J. Winicour${}^{3,5}$
       }
\affiliation{
${}^{1}$ Department of Physics \\
         Marshall University, Huntington, WV 25755, USA \\
${}^{2}$Department of Mathematical Sciences,
          University of South Africa, Unisa 0003, South Africa,\\
${}^{3}$ Max-Planck-Institut f\" ur
         Gravitationsphysik, Albert-Einstein-Institut, 
	 14476 Golm, Germany \\
${}^{4}$ Theoretical Astrophysics, California Institute of Technology \\
               Pasadena, CA 91125, USA \\
${}^{5}$ Department of Physics and Astronomy \\
        University of Pittsburgh, Pittsburgh, PA 15260, USA
	 }

\begin{abstract}

We develop, test and compare new numerical and geometrical methods for improving
the accuracy of extracting waveforms using characteristic evolution. The new
numerical method involves use of circular boundaries to the stereographic grid
patches which cover the spherical cross-sections of the outgoing null cones. We
show how an angular version of numerical dissipation can be introduced into the
characteristic code to damp the high frequency error arising form the irregular
way the circular patch boundary cuts through the grid. The new geometric method
involves use of the Weyl tensor component $\Psi_4$ to extract the waveform as
opposed to the original approach via the Bondi news function. We develop the
necessary analytic and computational formula to compute the $O(1/r)$
radiative part of $\Psi_4$ in terms of a conformally compactified treatment of
null infinity. These methods are compared and calibrated in test problems based
upon linearized waves.

\end{abstract}


\maketitle

\section{Introduction}

The unambiguous geometric description of gravitational radiation in curved
spacetimes traces back to the work of Bondi~\cite{bondi} et al,
Sachs~\cite{sachs} and Penrose~\cite{Penrose}. By formulating asymptotic
flatness in terms of characteristic hypersurfaces extending to infinity, they
were able to reconstruct, in a nonlinear geometric setting, the basic properties
of gravitational waves which had been developed in linearized theory on a
Minkowski background. The major new nonlinear features were the Bondi mass and
news function, and the mass loss formula relating them. This approach has been
implemented as a characteristic evolution code~\cite{isaac,highp} which computes
the radiation field at infinity by using a Penrose compactification of the
space-time. The code computes the gravitational radiation reaching infinity in
terms of  boundary data supplied on an inner worldtube. This has timely
application to the important astrophysical problem of the inspiral and merger of
a binary black hole. Several Cauchy codes, using an artificial outer boundary
condition, are now able to simulate this binary problem. By using the data on a
worldtube carved out of these binary black hole spacetimes obtained by Cauchy
evolution, the characteristic code can supply the resulting waveform at
infinity. In this work, we develop and test new methods designed to enhance the
accuracy of this approach to computing gravitational waveforms, which has been
called {\it Cauchy-characteristic extraction} (CCE)~\cite{cce}.

The Cauchy codes presently being applied to the binary black hole problem
introduce an artificial outer boundary, where some boundary condition must be
employed. The choice of the proper boundary condition for an isolated radiating
system is a global problem, which can only be treated exactly by an extension of
the solution to infinity, e.g.  by conformal compactification. The most elegant such
approach is the extension of the Cauchy problem to future null infinity ${\cal
I}^+$ by means of a hyperboloidal time foliation~\cite{hypb}
(see~\cite{Hhprbloid,joerg} for reviews of progress in this direction). Another
approach is to extend the solution to  ${\cal I}^+$ by matching the interior
Cauchy evolution to an exterior characteristic evolution, i.e.
Cauchy-characteristic matching (CCM)~\cite{ccm}. (see~\cite{livccm} for a review). CCM has
been applied successfully to gravitational wave computations in the linear
regime~\cite{harm} but has not yet been extended to the nonlinear binary black
hole problem.

When an artificial finite outer boundary is introduced there are two broad
sources of error:
\begin{itemize}

\item The outer boundary condition

\item Waveform extraction at an inner worldtube.
\end{itemize}

The first source of error stems from the outer boundary condition, which must
lead to a well-posed constraint-preserving initial-boundary value problem. This
has not yet been fully established for any of the present black hole codes. But,
even were such boundary conditions implemented, the correct boundary data must
be prescribed. However, this boundary data can only be exactly determined, in
general, by extending the solution to infinity~\cite{gk}.  Otherwise, the best
that can be done is to impose a boundary condition for which {\it homogeneous}
boundary data, i.e. zero boundary values, is a good approximation. One proposal
of this type~\cite{friednag} is a boundary condition that requires the
Newman-Penrose~\cite{NP} Weyl tensor component $\Psi_0$ to vanish. In the limit that
the outer boundary goes to infinity this outer boundary condition becomes exact.
In the present state of the art of black hole simulations, this approach comes
closest to a satisfactory treatment of the outer boundary~\cite{psiocit}.

The second source of error arises from waveform extraction at an inner
worldtube, which must be well inside the outer boundary in order to isolate it
from errors introduced by the boundary condition. There the waveform is
typically extracted by a perturbative scheme based upon the introduction of a
background Schwarzschild spacetime. This has been carried out using the 
Regge-Wheeler-Zerilli ~\cite{regwh,zeril} treatment of the perturbed metric, as
reviewed in~\cite{nagrez},  and also by calculating the Newman-Penrose Weyl
component $\Psi_4$, as first done for the binary black hole problem
in~\cite{bakcamluo,pretlet,camluomarzlo,bakcenchokopvme}. In this approach,
errors arise from the finite size of the extraction worldtube, from
nonlinearities and from gauge ambiguities involved in the arbitrary introduction
of a background metric. The gauge ambiguities might seem less severe in the case
of $\Psi_4$ (vs metric) extraction, but there are still delicate problems
associated with the choices of a preferred null tetrad and  preferred worldlines
along which to measure the waveform (see ~\cite{lehnmor} for an analysis).

Cauchy-characteristic extraction, which is one of the pieces of the CCM
strategy, offers a means to avoid this error introduced by extraction at a
finite worldtube. In CCE, the inner worldtube data supplied by the Cauchy
evolution is used as boundary data for a characteristic evolution to future null
infinity, where the waveform can be unambiguously computed by geometric methods.
By itself, CCE does not use the characteristic evolution to inject outer
boundary data for the Cauchy evolution, which can be a source of instability in
full CCM. Highly nonlinear tests in black hole spacetimes~\cite{highp} have
shown that characteristic evolution is a stable procedure which provides the
geometry in the neighborhood of null infinity up to numerical error; and tests
in the perturbative regime~\cite{babiuc05} show that CCE compares favorably with
Zerilli extraction and has advantages at small extraction radii. However, in
astrophysically realistic cases which require high resolution, such as the
inspiral of matter into a black hole~\cite{partbh}, this error has been a
troublesome factor in the postprocessing of the numerical solution which is
necessary to compute the asymptotic quantities determining the Bondi news
function.

There are two distinct ways, geometric and numerical, that the accuracy of this
calculation of the gravitational waveform might be improved. In the geometrical
category, one option is to  compute $\Psi_4$ instead of the news function as the
primary description of the waveform. We discuss this in
Sec.~\ref{sec:waveforms}, where we develop the extensive formulae necessary  to
compute the asymptotic $O(1/r)$ part of $\Psi_4$, i.e.
$\Psi_4^0=\lim_{ r\rightarrow \infty }r\Psi_4$, which governs the radiation.  

In the numerical category, some standard methods for improving accuracy, such
as  higher order finite difference approximations, would be easy to implement
whereas others, such as adaptive mesh refinement, have only been tackled for 1D
characteristic codes~\cite{pretlehn}. But beyond these methods, a major source
of error in characteristic evolution arises from the intergrid interpolations
arising from the multiple patches necessary to coordinatize smoothly the
spherical cross-sections of the outgoing null hypersurfaces. The development of
grids smoothly covering the sphere has had a long history in computational
meteorology that has led to two distinct approaches: (i) the stereographic
approach in which the sphere is covered by two overlapping patches obtained by
stereographic projection about the North and South poles~\cite{browning}; and
(ii) the cubed-sphere approach in which the sphere is covered by the 6 patches
obtained by a projection of the faces of a circumscribed cube~\cite{ronchi}. 
Recently, the cubed-sphere approach has received much attention because the
simple structure of its shared boundaries allows a highly scalable algorithm for
parallel architectures. A discussion of the advantages of each of these methods
and a comparison of their performance in a standard fluid testbed are given
in~\cite{browning}. In numerical relativity, the stereographic method has been
reinvented in the context of the characteristic evolution problem~\cite{eth};
and the cubed-sphere method reinvented in the context of building an apparent
horizon finder~\cite{Thornburgah}. The characteristic evolution code was first
developed using two square stereographic patches, each overlapping the equator.
We consider here a modification, based  upon the approach advocated
in~\cite{browning}, which retains the original stereographic patch structure but
shrinks the overlap region by masking a circular boundary near the equator.
Recently, the cubed-sphere method has also been developed for application to
characteristic evolution~\cite{reisswig,roberto}.

These geometric and numerical considerations lead to four options for improving
CCE:

\begin{itemize}
 
\item Computation of the news function using circular stereographic patches 

\item Computation of the Weyl tensor using circular stereographic patches 

\item Computation of the news function using the cubed-sphere patching 

\item Computation of the Weyl tensor using cubed-sphere patching 

\end{itemize}

We compare these options here in the context of model problems designed to test
their application to CCE. Because the cubed-sphere approach requires further
code development to be applied to CCE, in Sec.~\ref{sec:stests} we present a
test based upon the propagation of a wave on the sphere to provide a preliminary
comparison with the stereographic approach. The test  compares their accuracy in
calculating the angular derivatives required in the news and Weyl tensor
extraction algorithms. In Sec.~\ref{sec:extraction}, we next present tests of
CCE which compare the news and Weyl tensor approaches in a linearized
gravitational wave test problem.

The development of finite-difference evolution algorithms, which was largely
motivated by application to computational fluid dynamics (CFD). It has utilized
the method of lines, where a 3-dimensional spatial domain is discretized to
yield a set of coupled ordinary differential equations in time for the grid
values, which are then integrated, e.g. by a Runge-Kutta procedure. This $3+1$
approach is not applicable to the $2+1+1$ format of characteristic evolution
considered here, in which the discretization of a 2-dimensional spherical set of
characteristics leads to coupled 2-dimensional partial differential equations in
the plane spanned by the outgoing and ingoing characteristics. This $2+1+1$
approach is natural to general relativity since the characteristics (light rays)
are fundamental to describing the dynamical geometry of space-time. It would be
impractical in CFD in which the characteristics have a complicated dynamic
relation (determined by equations of state) to the fixed Euclidean geometry. As
a result, characteristic evolution algorithms were developed only recently in the
context of general relativity and there has been relatively little analysis of
their computational properties. In particular, for CFD or any symmetric
hyperbolic system, numerical dissipation can be added in the standard
Kreiss-Oliger form~\cite{kreisoligd}. One of the main results of this paper is
to show how analogous dissipation can be successfully applied in a $2+1+1$
format. In the original version of the PITT code, which used square
stereographic patches with boundaries aligned with the grid, numerical
dissipation was only introduced in the radial direction~\cite{luisdis}. This was
sufficient to establish numerical stability. In the new version of the code with
circular stereographic patches, whose boundaries do not fit regularly on the
stereographic grid, numerical dissipation is necessary to control the high
frequency error introduced by the intergrid interpolations, as previously noted
in the treatment of a fluid problem using circular stereographic
patches~\cite{browning}. We begin with a brief review of the formalism
underlying the characteristic evolution code in Sec.~\ref{sec:chform} and show
how the essential new feature of angular dissipation can be incorporated. 

The two spherical grid methods, stereographic and cubed sphere, are briefly
described in Sec.~\ref{sec:patches}. We present the test results in
Secs.~\ref{sec:stests} and \ref{sec:extraction} and we summarize our conclusions
in  Sec.~\ref{sec:concl}. 

\section{Characteristic Formalism}
\label{sec:chform}

The characteristic formalism is based upon a family of outgoing null
hypersurfaces, emanating from some inner worldtube, which extend to
infinity where they foliate ${\cal I}^+$ into spherical slices. 
We let $u$ label these hypersurfaces, $x^A$ $(A=2,3)$ be angular coordinates
which label the null rays and $r$ be a surface area coordinate. In the
resulting $x^\alpha=(u,r,x^A)$ coordinates, the metric takes the Bondi-Sachs
form~\cite{bondi,sachs}
\begin{eqnarray}
   ds^2 & = & -\left(e^{2\beta}\frac{V}{r} -r^2h_{AB}U^AU^B\right)du^2
        -2e^{2\beta}dudr -2r^2 h_{AB}U^Bdudx^A \nonumber \\
        & + & r^2h_{AB}dx^Adx^B,    
	\label{eq:bmet}
\end{eqnarray}
where $h^{AB}h_{BC}=\delta^A_C$ and
$det(h_{AB})=det(q_{AB})$, with $q_{AB}$ a unit sphere metric.  In
analyzing the Einstein equations, we also use the intermediate variable
\begin{equation}
     Q_A = r^2 e^{-2\,\beta} h_{AB} U^B_{,r}.
\end{equation}

The code introduces an auxiliary unit sphere metric $q_{AB}$, with
associated complex dyad $q_A$ satisfying
$ q_{AB} =\frac{1}{2}\left(q_A \bar q_B+\bar q_Aq_B\right)$.
For a general Bondi-Sachs metric,
$h_{AB}$ can then be represented by its dyad component $J=h_{AB}q^Aq^B
/2$, with the spherically symmetric case characterized by $J=0$. The
full nonlinear $h_{AB}$ is uniquely determined by $J$, since the
determinant condition implies that the remaining dyad component
$K=h_{AB}q^A \bar q^B /2$ satisfies $1=K^2-J\bar J$. We also introduce
spin-weighted fields $U=U^Aq_A$ and $Q=Q_Aq^A$, as well as the (complex
differential) operators $\eth$ and $\bar \eth$~\cite{newp}.
Refer to {}~\cite{eth,cce} for further details.

In this formalism, the Einstein equations $G_{\mu\nu}=0$ decompose into
hypersurface equations, evolution equations and conservation conditions on the
inner worldtube. As described in more detail in~\cite{newt,nullinf}, the
hypersurface equations take the form
\begin{eqnarray}
      \beta_{,r} &=& N_\beta, 
   \label{eq:beta} \\
          U_{,r}  &=& r^{-2}e^{2\beta}Q +N_U, 
     \label{eq:wua} \\
     (r^2 Q)_{,r}  &=& -r^2 (\bar \eth J + \eth K)_{,r}
                +2r^4\eth \left(r^{-2}\beta\right)_{,r} + N_Q, 
     \label{eq:wq} \\
V_{,r} &=& \frac{1}{2} e^{2\beta}{\cal R} 
- e^{\beta} \eth \bar \eth e^{\beta}
+ \frac{1}{4} r^{-2} \left(r^4
                           \left(\eth \bar U +\bar \eth U \right)
                     \right)_{,r} + N_W,
 \label{eq:ww}
\end{eqnarray}
where~\cite{eth}
\begin{equation}
{\cal R} =2 K - \eth \bar \eth K + \frac{1}{2}(\bar \eth^2 J + \eth^2 \bar J)
          +\frac{1}{4K}(\bar \eth \bar J \eth J - \bar \eth J \eth \bar J)
     \label{eq:calR}
\end{equation}
is the curvature scalar of the 2-metric $h_{AB}$.
The evolution equation takes the form
\begin{eqnarray}
    && 2 \left(rJ\right)_{,ur}
    - \left(r^{-1}V\left(rJ\right)_{,r}\right)_{,r} = \nonumber \\
    && -r^{-1} \left(r^2\eth U\right)_{,r}
    + 2 r^{-1} e^{\beta} \eth^2 e^{\beta}- \left(r^{-1} V \right)_{,r} J
    + N_J,
    \label{eq:wev}
\end{eqnarray}
where, $N_\beta$, $N_U$, $N_Q$, $N_W$ and $N_J$ are nonlinear terms  which
vanish for spherical symmetry. Expressions for these terms as complex
spin-weighted fields and a discussion of the conservation conditions are given
in~\cite{cce}.

The characteristic evolution code implements this formalism as an explicit
finite difference scheme. In this paper we use second order accurate finite
differences and we reduce all angular derivatives to first order by the
introduction of auxiliary variables, as described in~\cite{gomezfo}

\subsection{Angular dissipation}
\label{sec:angdiss}

It is a feature of the composite mesh technique that numerical dissipation is
necessary to stabilize the error introduced by intergrid interpolations. In the
case of a square stereographic patch, whose boundary aligns with the grid lines,
the dissipation built into the characteristic radial integration scheme is
sufficient  for this purpose~\cite{luisdis}. However, because a circular
boundary fits into a stereographic grid in an irregular way,  angular
dissipation is also necessary in order to suppress the resulting high frequency
error introduced by the interpolations between stereographic patches.

We accomplish this by modifying the evolution equation (\ref{eq:wev}) as follows. In the
code, (\ref{eq:wev}) is expressed in terms of a compactified radial coordinate
$x=r/(R+r)$, where $R$ is an adjustable scale parameter and ${\cal I}^+$ has
finite coordinate value $x=1$. The evolution in retarded time $u$ is carried out in terms of the variable
$\Phi=xJ$, which is regular at ${\cal I}^+$. Then the evolution equation
(\ref{eq:wev}) takes the form 
\begin{equation}
   \partial_u \bigg((1-x)  \Phi_{,x} +\Phi \bigg)= S,
\label{eq:phiev}
\end{equation}
where $S$ represents the right hand side terms.
We add angular dissipation to the $u$-evolution through the modification
\begin{equation}
    \partial_u \bigg ((1-x)  \Phi_{,x} +\Phi \bigg )   
  +\epsilon_{u} h^3 \eth^2 {\cal W} 
      \bar\eth^2 \bigg ((1-x)  \Phi_{,x} +\Phi \bigg ) 
      = S,
\end{equation}
where $h$ is the discretization size, $\epsilon_{u}\ge 0$ is an adjustable
parameter independent of $h$ and $ {\cal W}$ is a positive weighting function
with $ {\cal W}=1$ inside the equator and $ {\cal W}=0$ at the patch boundary.
This leads to 
\begin{equation} 
     \partial_u \bigg( |(1-x)  \Phi_{,x} +\Phi|^2\bigg) 
     + 2\epsilon_{u}  h^3 \Re\{ \bigg ((1-x) \bar \Phi_{,x} +\bar \Phi \bigg ) 
       \eth^2  {\cal W}\bar\eth^2 \bigg ((1-x)  \Phi_{,x} +\Phi \bigg )    \} 
      = 2\Re \{\bigg ((1-x) \bar \Phi_{,x} +\bar \Phi \bigg ) S \}.
\end{equation}
Integration over the unit sphere with solid angle element $d\Omega$ then gives
\begin{equation} 
   \partial_u \oint | (1-x) \Phi_{,x} +\Phi|^2  d\Omega 
     +2\epsilon_{u}  h^3 \oint 
            {\cal W}|\bar\eth^2 \bigg ((1-x)\Phi_{,x}+\Phi \bigg )|^2 d\Omega 
         =2\Re \oint \bigg ((1-x) \bar \Phi_{,x} +\bar \Phi \bigg ) S d\Omega.
\end{equation}
Thus the $\epsilon_{u}$-term has the effect of damping high frequency noise as
measured by the $L_2$ norm of $(1-x) \Phi_{,x} +\Phi$ over the sphere. 

Similarly, dissipation is introduced in the radial integration of
(\ref{eq:phiev}) through the substitution
\begin{equation}
    \partial_u \bigg ((1-x)  \Phi_{,x} +\Phi \bigg )  \rightarrow 
 \partial_u \bigg ((1-x)  \Phi_{,x} +\Phi \bigg ) 
  +\epsilon_{x} h^3 \eth^2  {\cal W} \bar\eth^2  \Phi_{,u} 
\end{equation}
with $\epsilon_{x}\ge 0$ .
Angular dissipation is also introduced in the hypersurface
equations  through the substitutions
\begin{eqnarray}
          (r^2 Q)_{,r}  &\rightarrow& (r^2 Q)_{,r}
                +\epsilon_Q h^3 \eth \bar \eth  {\cal W} \eth\bar \eth r Q      \\
          V_{,r} &\rightarrow & V_{,r} 
                +\epsilon_V h^3 \eth\bar\eth  {\cal W}  \eth\bar\eth V.
\end{eqnarray}

\section{Waveforms at ${\cal I}^+$}
\label{sec:waveforms} 

For an analytic treatment of the Penrose compactification of an asymptotically flat
space-time, it is simplest to introduce an inverse radial coordinate $\ell=1/r$,
so that future null infinity ${\cal I}^+$ is given by $\ell=0$~\cite{tam}. In
the resulting $x^\mu=(u,\ell,x^A)$ conformal Bondi coordinates, the physical
space-time metric $g_{\mu\nu}$ has the conformal compactification $\hat
g_{\mu\nu}=\ell^{2} g_{\mu\nu}$, where $\hat g_{\mu\nu}$ is smooth at ${\cal
I}^+$ and, referring to (\ref{eq:bmet}), takes the form
\begin{equation}
   \hat g_{\mu\nu}dx^\mu dx^\nu= 
           -\left(e^{2\beta}V \ell^3 -h_{AB}U^AU^B\right)du^2
        +2e^{2\beta}dud\ell -2 h_{AB}U^Bdudx^A + h_{AB}dx^Adx^B.
   \label{eq:lmet}
\end{equation}
The inverse conformal metric has the non vanishing components
$\hat g^{u\ell} =e^{-2\beta}$,
$\hat g^{\ell \ell} =e^{-2\beta}\ell^3 V$, 
$\hat g^{\ell A} =e^{-2\beta}U^A$ and $\hat g^{AB}=h^{AB}$.

The Bondi mass, news function and $\Psi_4^0$ (functions of $u$ and  $x^A$),
which describe the total energy and radiation power, are constructed from the
leading coefficients in an expansion of the metric in powers of $\ell$. The
requirement of an asymptotically flat vacuum exterior imposes relations between
these expansion coefficients. In the $\hat g_{\mu\nu}$ conformal frame,  the
vacuum gravitational equations are 
\begin{equation}
     -\ell^2 \hat G_{\mu\nu} =2\ell (\hat\nabla_\mu \hat\nabla_\nu \ell
           -\hat g_{\mu\nu} \hat \nabla^\alpha \hat\nabla_\alpha \ell ) 
          +3\hat g_{\mu\nu} (\hat\nabla^\alpha \ell) \hat\nabla_\alpha \ell 
	  \label{eq:einstein}
\end{equation}
in terms of the Einstein tensor $\hat G_{\mu\nu}$ and covariant derivative $\hat
\nabla_\mu$ associated with $\hat g_{\mu\nu}$. Asymptotic flatness immediately
implies that $\hat g^{\ell \ell} =(\hat\nabla^\alpha \ell) \hat\nabla_\alpha
\ell =O(\ell)$ so that ${\cal I}^+$ is null. From the trace of
(\ref{eq:einstein}), we have
\begin{equation}
   (\hat\nabla^\alpha \ell) \hat\nabla_\alpha \ell= 
             \frac{1}{2}\ell \hat\Theta+O(\ell^2),
\end{equation}
where 
\begin{equation}
   \hat\Theta:=\hat\nabla^\mu \hat\nabla_\mu \ell 
   =e^{-2\beta} \bigg( \partial_\ell (\ell^3 V)+\partial_A U^A \bigg)
\label{eq:theta}
\end{equation}
is smooth at ${\cal I}^+$. In addition, (\ref{eq:einstein}) implies the
existence of a smooth trace-free field $\hat\Sigma_{\mu\nu}$ defined by
\begin{equation}
     \ell \hat \Sigma_{\mu\nu} := 
      \hat \nabla_\mu \hat\nabla_\nu \ell
          -\frac{1}{4}\hat g_{\mu\nu}\hat\Theta. 
\label{eq:Sigma}
\end{equation}
For future reference we introduce an orthonormal null tetrad $(\hat n^\mu, \hat
\ell^\mu, \hat m^\mu)$ be  such that $\hat n^\mu=\hat \nabla^\mu \ell$ and 
$\hat \ell^\mu \partial\mu=\partial_\ell$ at ${\cal I}^+$. Note that
(\ref{eq:einstein}), (\ref{eq:theta}) and (\ref{eq:Sigma}) imply
\begin{equation}
  \hat m^\nu \hat m^\rho (\hat\Sigma_{\nu\rho}+\frac{1}{2}\hat G_{\nu\rho})=0.
\label{eq:einsteinsig}
\end{equation}
The gravitational waveform depends on the value of $\hat\Sigma_{\mu\nu}$ on
${\cal I}^+$, which in turn depends on the leading terms up to $O(\ell)$ in the
expansion of $\hat g_{\mu\nu}$. We thus expand
\begin{equation}
   h_{AB}(u,\ell,x^C)= H_{AB}(u,x^C)+\ell c_{AB}(u,x^C)+O(\ell^2).
\end{equation}
Further conditions on the asymptotic expansion of the metric can be extracted
from (\ref{eq:einstein}). We have
\begin{equation}
    \beta(u,\ell,x^C)=H(u,x^C)+ O(\ell^2) 
\end{equation}
(where the $O(\ell)$ term vanishes),
\begin{equation}
    U^A= L^A+2\ell e^{2H} H^{AB}D_B H+O(\ell^2),
\end{equation}
and 
\begin{equation}
    \ell^2 V= D_A L^A
     +\ell (e^{2H}{\cal R}/2 +D_A D^A e^{2H})+O(\ell^2),
\end{equation}
where ${\cal R}$ and $D_A$ are the 2-dimensional curvature scalar and covariant
derivative associated with $h_{AB}$. These results combine with (\ref{eq:theta})
to give 
\begin{equation}
     \hat\Theta =2e^{-2H}D_A L^A +\ell \bigg ({\cal R}
           + 3 e^{-2H}D_A D^A e^{2H} \bigg )+O(\ell^2).     
\end{equation}
In addition, the requirement that
$$
     \ell ( \hat \Sigma_{AB} -\frac{1}{2}H_{AB} H^{CD}\Sigma_{CD})
$$
vanishes at ${\cal I}^+$ implies via (\ref{eq:Sigma}) that
\begin{equation}
     2H_{C (A} D_{B)} L^C+\partial_u H_{AB}
        - H_{AB}  D_{C} L^C =O(\ell).
\label{eq:asymHu}
\end{equation}

The expansion coefficients $H$, $H_{AB}$, $c_{AB}$ and $L^A$ (all functions of
$(u,x^A)$) completely determine the radiation field. One can further specialize
the Bondi coordinates to be {\em inertial} at ${\cal I}^+$, i.e. have Minkowski
form, in which case $H=L^A=0$, $H_{AB}=q_{AB}$ (the unit sphere metric) so that
(\ref{eq:asymHu}) is trivially satisfied and the radiation field is determined
by $c_{AB}$. However, the characteristic extraction of the waveform is carried
out in null coordinates determined by data on the inner worldtube so that this
{\em inertial} simplification cannot be assumed.

\subsection{Calculation of the news}

The following calculation of the Bondi news streamlines the presentation
in~\cite{highp} and corrects errors. In order to carry out the
calculation in the $\hat g_{\mu\nu}$ computational frame, it is useful to refer
to an inertial conformal Bondi frame~\cite{tam} with metric $\tilde
g_{\mu\nu}=\Omega^2 g_{\mu\nu} =\omega^2 \hat g_{\mu\nu}$, where
$\Omega=\omega\ell$, which satisfies the gauge requirements that
$Q_{AB}:={\tilde g}_{AB}|_{{\cal I}^+}=\omega^2 H_{AB}$ is intrinsically a unit
sphere metric at ${\cal I^+}$ and that $(\tilde \nabla^{\alpha} \Omega)\tilde
\nabla_{\alpha} \Omega=O(\Omega^2)$. (See~\cite{quad} for a discussion of how
the news in an arbitrary conformal frame is related to its expression in this
inertial Bondi frame.)

${\cal I^+}$ is a null hypersurface with the null vector ${\tilde
n}^{\alpha}={\tilde g}^{\alpha \beta}\nabla_{\beta}  \Omega |_{\cal I^+}$, or
equivalently,  ${\hat n}^{\alpha} = {\hat g}^{\alpha\beta}\nabla_{\beta} \ell
|_{\cal I^+}=\omega{\tilde n}^{\alpha}$, tangent to its generators. In order to
complete a basis for tangent vectors to ${\cal I^+}$, let ${\cal Q}^{\alpha}$ be
a complex field tangent to ${\cal I^+}$ satisfying ${\cal Q}^{\alpha} \tilde
n_\alpha=0$, ${\tilde g}_{\alpha\beta}{\cal Q}^{\alpha}{\cal Q}^{\beta} |_{\cal
I^+}=0$ and ${\tilde g}_{\alpha\beta}{\cal Q}^{\alpha}{\bar {\cal Q}}^{\beta}
|_{\cal I^+}=2$. In an inertial conformal Bondi frame, the news function can
then be expressed as~\cite{highp}
\begin{equation}
 N=\lim_{\Omega \rightarrow 0}{1\over 2\Omega}{\cal Q}^{\alpha} {\cal Q}^{\beta}
     {\tilde \nabla}_{\alpha} {\tilde \nabla}_{\beta}  \Omega 
\label{eq:snews}
\end{equation}
evaluated in the limit of ${\cal I^+}$. (Our conventions are chosen so
that the news reduces to Bondi's original expression in
the axisymmetric case~\cite{bondi}). In terms of
the ${\hat g}_{\alpha\beta}$ frame, with conformal factor $ \ell=\Omega/\omega$,
we then have
\begin{eqnarray}
 N=\lim_{\ell \rightarrow 0}{1\over 2}{\cal Q}^{\alpha} {\cal Q}^{\beta} \bigg(
  {{\hat \nabla}_{\alpha} {\hat \nabla}_{\beta}
          {\ell} \over {\ell}}
   -\omega{\hat \nabla}_{\alpha} {\hat \nabla}_{\beta} {1\over \omega}
     +\frac{1}{\ell \omega}\hat g_{\alpha\beta} 
        (\hat\nabla^\mu \ell) \hat\nabla_\mu \omega                  \bigg ) \\
       ={1\over 2}{\cal Q}^{\alpha} {\cal Q}^{\beta} \bigg (
  \hat \Sigma_{\alpha\beta}
   -\omega{\hat \nabla}_{\alpha} {\hat \nabla}_{\beta} {1\over \omega}
     +\frac{1}{\omega}(\partial_\ell \hat g_{\alpha\beta} )
         (\hat\nabla^\mu \ell) \hat\nabla_\mu \omega  
                 \bigg ).
     \label{eq:lnews}
\end{eqnarray}
(This corrects an error in equation (30) of~\cite{highp}.) We determine $\omega$
on ${\cal I}^+$ in the ${\hat g}_{\alpha\beta}$ frame by solving the elliptic
equation governing the conformal transformation of the curvature scalar
(\ref{eq:calR}) of the geometry intrinsic to a $u=constant$ cross-section
to a unit sphere geometry,
\begin{equation}
     {\cal R}=2(\omega^2+H^{AB}D_A D_B \log \omega).
\label{eq:conf}
\end{equation}
The condition that  $(\tilde \nabla^{\alpha} \Omega) \tilde \nabla_{\alpha}
\Omega=O(\Omega^2)$ determines the time dependence of $\omega$,
\begin{equation}
     2{\hat n}^{\alpha} \partial_{\alpha} \log \omega
    =-e^{-2H}D_AL^A,
\label{eq:omegadot}
\end{equation}
which is used to evolve $\omega$ given a solution of
(\ref{eq:conf}) as initial condition.

In order to obtain an explicit expression for the news (\ref{eq:lnews}) in the
${\hat g}_{\alpha\beta}$ frame we need to fix the choice of ${\cal Q}^{\beta}$.
The freedom ${\cal Q}^{\beta} \rightarrow {\cal Q}^{\beta} + \lambda {\tilde
n}^{\beta}$ leaves (\ref{eq:lnews}) invariant but it is important for physical
interpretation to choose the spin rotation freedom ${\cal Q}^{\beta} \rightarrow
e^{-i\alpha} {\cal Q}^{\beta}$ to satisfy ${\tilde n}^{\alpha}{\tilde
\nabla}_{\alpha} {\cal Q}^{\beta}=O(\Omega)$, so that the polarization frame is
parallel propagated along the generators of ${\cal I}^+$. This fixes the
polarization modes determined by the real and imaginary parts of the news to
correspond to those of inertial observers at ${\cal I}^+$. 

We accomplish this by introducing the dyad
decomposition  $H^{AB}=(F^A{\bar F}^B+{\bar F}^A  F^B)/2$ where
\begin{equation}
   F^A  = q^A  \sqrt{ \frac{(K+1)}{2 }  }
          -\bar q^A  J \sqrt{ 1 \over 2(K+1)} .
\end{equation}
We set ${\cal Q}^{\beta}=e^{-i\delta}\omega^{-1}F^\beta+\lambda {\tilde
n}^{\beta}$, where $F^\alpha:=(0,0,F^A)$. The requirement of an inertial
polarization frame,  ${\tilde n}^{\alpha}{\tilde \nabla}_{\alpha} {\cal
Q}^{\beta}=O(\Omega)$, then determines the time dependence of the phase
$\delta$. We obtain, after using (\ref{eq:omegadot}) to eliminate the time
derivative of $\omega$,
\begin{equation}
    2i(\partial_u +L^A\partial_A)\delta = D_A L^A
     +H_{AC} \bar F^C ( (\partial_u +L^B \partial_B) F^A
             - F^B \partial_B L^A) .
    \label{eq:evphase}
\end{equation}

We can now express the inertial news (\ref{eq:lnews}) in the ${\hat
g}_{\alpha\beta}$ frame as
\begin{equation}
  N={1\over 2}e^{-2i\delta}\omega^{-2}F^{\alpha} F^{\beta}  \bigg(
       \hat \Sigma_{\alpha\beta}
   -\omega{\hat \nabla}_{\alpha} {\hat \nabla}_{\beta} {1\over \omega}
     +\frac{1}{\omega}(\partial_\ell \hat g_{\alpha\beta} )
         (\hat\nabla^\mu \ell) \hat\nabla_\mu \omega             \bigg).
   \label{eq:hnews}
\end{equation}
with $F^{\alpha}=(0,0,F^A)$. An explicit calculation leads to
\begin{equation}
    N={1\over 4}e^{-2i \delta}\omega^{-2}e^{-2H}F^A F^B
       \{(\partial_u+{\pounds_L})c_{AB}-{1\over 2}c_{AB} D_C L^C
        +2\omega D_A[\omega^{-2}D_B(\omega e^{2H})]\},
     \label{eq:news}
\end{equation}
where $\pounds_L$ denotes the Lie derivative with respect to $L^A$. This
corrects a minus sign error in (38) of~\cite{highp}, where spin-weighted 
expressions for the terms in (\ref{eq:news}) are given.

In inertial Bondi coordinates, the expression for the news function reduces to
the simple form
\begin{equation}
    N={1\over 4}Q^A Q^B \partial_u c_{AB}.
     \label{eq:inews}
\end{equation}
However, the general form (\ref{eq:news}) must be used in the computational
coordinates, which is challenging for maintaining accuracy because of the
appearance of second angular derivatives of $\omega$. 

\subsection{Calculation of the Weyl tensor} \label{sec:weyl}

Asymptotic flatness implies that the Weyl tensor vanishes at ${\cal I}^+$, i.e.
$\hat C_{\mu\nu\rho\sigma}=O(\ell)$ in the $\hat g_{\mu\nu}$ conformal Bondi
frame (\ref{eq:lmet}). This is the conformal space version of the peeling
property of asymptotically flat spacetimes~\cite{Penrose}. In terms of the
orthonormal null tetrad $(\hat n^\mu, \hat \ell^\mu, \hat m^\mu)$, with
$\hat n^\mu=\hat \nabla^\mu \ell$ and  $\hat \ell^\mu \partial\mu=\partial_\ell$ at ${\cal I}^+$, the radiation is described by the limit 
\begin{equation}
      \hat \Psi:=-\frac{1}{2} \lim_{\ell \rightarrow 0}\frac{1}{\ell}
    \hat n^\mu \hat m^\nu \hat n^\rho \hat m^\sigma \hat C_{\mu\nu\rho\sigma},
\label{eq:psi}
\end{equation}
which corresponds in Newman-Penrose notation to $-(1/2)\bar \psi_4^0$. The
limit is independent of how the tetrad is extended off ${\cal I}^+$ but
to simplify the calculation we make the following choices adapted to
our conformal Bondi coordinates. We set
$\hat \ell^\mu =e^{2\beta}\hat \nabla^\mu u$,
$\hat n^\mu =\hat \nabla^\mu \ell +O(\ell)$,
$\hat \ell^\rho \hat \nabla_\rho \hat m^\mu=0$ and
$\hat \ell^\rho \hat \nabla_\rho \hat n^\mu=0$,
which implies
\begin{equation}
      \hat n_\mu= \hat \nabla_\mu \ell
       - \frac{\ell}{4}\hat \ell_\mu \hat\Theta +O(\ell^2).
\label{eq:hatn} 
\end{equation}

Our main calculational result is:
\begin{equation}
       \hat \Psi=\frac{1}{2}\hat n^\mu \hat m^\nu \hat m^\rho \bigg(
	         \hat \nabla_\mu  \hat \Sigma_{\nu\rho}
		  -\hat \nabla_\nu \hat \Sigma_{\mu\rho}\bigg)|_{\cal I^+} ,
\label{eq:psisigma}
\end{equation}
and that (\ref{eq:psisigma}) is independent of the freedom
\begin{equation}
      \hat m^\nu \rightarrow \hat m^\nu +\lambda \hat n^\nu.	          
\label{eq:mnfreedom}
\end{equation}

The result (\ref{eq:psisigma}) follows from the following sequence of
calculations beginning with (\ref{eq:psi}) (where evaluation at  ${\cal I}^+$ is
assumed):
\begin{eqnarray}
     -2 \hat \Psi &=&\frac{1}{\ell}
          \hat n^\mu \hat m^\nu \hat n^\rho \hat m^\sigma 
	  \hat C_{\mu\nu\rho\sigma} 
\label{eq:stepi}\\
	  &=&\frac{1}{\ell}
	  \hat n^\mu \hat m^\nu \hat n^\rho \hat m^\sigma 
	  \hat R_{\mu\nu\rho\sigma} 
\label{eq:trace}\\
	  &=&-\frac{1}{\ell}\hat n^\mu \hat m^\nu \hat m^\rho 
	  (\hat \nabla_\mu\hat \nabla_\nu \hat n_\rho 
	  -\hat \nabla_\nu \hat \nabla_\mu \hat n_\rho ) 
\label{eq:commut} \\
	  &=&-\frac{1}{\ell}\hat n^\mu \hat m^\nu \hat m^\rho 
	  \bigg(
	         \hat  \nabla_\mu (\ell \hat \Sigma_{\nu\rho}) 
	     -\hat \nabla_\nu (\ell \hat \Sigma_{\mu\rho}) 
	  - \hat \nabla_\mu\hat \nabla_\nu (\frac{\ell\hat\Theta}{4}\ell_\rho) 
	   +\hat \nabla_\nu\hat \nabla_\mu (\frac{\ell\hat\Theta}{4}\ell_\rho)
	     	   \bigg  ) 
\label{eq:sigma}\\
	   &=&-\hat n^\mu \hat m^\nu \hat m^\rho \bigg(
	         \hat \nabla_\mu  \hat \Sigma_{\nu\rho} 
	     -\hat \nabla_\nu \hat \Sigma_{\mu\rho} \bigg ) \nonumber \\
	   &-&\frac{1}{\ell}\hat n^\mu \hat m^\nu \hat m^\rho  \bigg(
	   \hat \Sigma_{\nu\rho} \hat \nabla_\mu \ell 
	    -\hat \Sigma_{\mu\rho} \hat \nabla_\nu \ell
	  - \hat \nabla_\mu\hat \nabla_\nu (\frac{\ell\hat\Theta}{4}\ell_\rho) 
	  +\hat \nabla_\nu\hat \nabla_\mu (\frac{\ell\hat\Theta}{4}\ell_\rho)
	     	   \bigg  )
\label{eq:diff}\\
	   &=&-\hat n^\mu \hat m^\nu \hat m^\rho \bigg(
	         \hat \nabla_\mu  \hat \Sigma_{\nu\rho} 
	     -\hat \nabla_\nu \hat \Sigma_{\mu\rho} \bigg ) \nonumber \\
	   &-&\frac{1}{\ell}\hat n^\mu \hat m^\nu \hat m^\rho  \bigg(
	    \hat \Sigma_{\nu\rho} \frac{\ell\hat\Theta}{4}\ell_\mu  
	    -\hat \Sigma_{\mu\rho}\frac{\ell\hat\Theta}{4}\ell_\nu  
	  - \hat \nabla_\mu\hat \nabla_\nu (\frac{\ell\hat\Theta}{4}\ell_\rho) 
	  +\hat \nabla_\nu\hat \nabla_\mu (\frac{\ell\hat\Theta}{4}\ell_\rho)
	     	   \bigg  ) 
\label{eq:hatn1}\\
           &=&-\hat n^\mu \hat m^\nu \hat m^\rho \bigg(
	         \hat \nabla_\mu  \hat \Sigma_{\nu\rho} 
	     -\hat \nabla_\nu \hat \Sigma_{\mu\rho} \bigg ) \nonumber \\
	   &-&\frac{\hat\Theta}{4} \hat m^\nu \hat m^\rho  \bigg (
	   \hat \Sigma_{\nu\rho}-n^\mu\ell^\sigma \hat R_{\mu\nu\rho\sigma}
	    \bigg ) 
\label{eq:lead}\\		   
            &=&-\hat n^\mu \hat m^\nu \hat m^\rho \bigg(
	         \hat \nabla_\mu  \hat \Sigma_{\nu\rho} 
	     -\hat \nabla_\nu \hat \Sigma_{\mu\rho} \bigg ) \nonumber \\
	   &-&\frac{\hat\Theta}{4} \hat m^\nu \hat m^\rho  \bigg (
	  \hat \Sigma_{\nu\rho}-\hat n^\mu \ell^\sigma 
	   (\hat g_{\mu[\nu}\hat G_{\sigma]\rho}
	          -\hat g_{\rho[\nu}\hat G_{\sigma]\mu}) \bigg )
\label{eq:vac} \\
		 &=&-\hat n^\mu \hat m^\nu \hat m^\rho \bigg(
	         \hat \nabla_\mu  \hat \Sigma_{\nu\rho} 
	     -\hat \nabla_\nu \hat \Sigma_{\mu\rho} \bigg ) \nonumber \\
	   &-&\frac{\hat\Theta}{4} \hat m^\nu \hat m^\rho  \bigg (
	   \hat\Sigma_{\nu\rho}+\frac{1}{2}\hat G_{\nu\rho})
\label{eq:algebra} \\	 
	   &=&-\hat n^\mu \hat m^\nu \hat m^\rho \bigg(
	         \hat \nabla_\mu  \hat \Sigma_{\nu\rho}
		  -\hat \nabla_\nu \hat \Sigma_{\mu\rho}\bigg)
\label{eq:stepf}. 
\end{eqnarray}
Here (\ref{eq:trace}) follows because all trace terms vanish; (\ref{eq:commut})
follows from the commutator of covariant derivatives; (\ref{eq:sigma}) follows
from (\ref{eq:Sigma}); (\ref{eq:diff}) follows from differentiation; 
(\ref{eq:hatn1}) follows from (\ref{eq:hatn}); (\ref{eq:lead}) follows from
taking leading terms and using the covariant commutator; (\ref{eq:vac}) follows
from the vanishing of the Weyl tensor at ${\cal I}^+$;  (\ref{eq:algebra})
follows algebraically; and  (\ref{eq:stepf}) follows from (\ref{eq:einsteinsig}).

Invariance of $\hat \Psi$ under the freedom (\ref{eq:mnfreedom}) follows from
noting that
\begin{equation}
 \hat n^\mu \hat m^\nu \hat n^\rho \hat n^\sigma 
	  \hat C_{\mu\nu\rho\sigma} =0
\end{equation}
and then following the steps analogous to (\ref{eq:stepi}) - (\ref{eq:stepf}) to
show
\begin{equation} 
        \hat n^\mu \hat m^\nu \hat n^\rho \bigg(
	         \hat \nabla_\mu  \hat \Sigma_{\nu\rho}
		  -\hat \nabla_\nu \hat \Sigma_{\mu\rho}\bigg)|_{\cal I^+}=0.
\end{equation}

Finally, the Weyl tensor must be scaled intrinsic to the $\tilde g_{\mu\nu}$
conformal frame in order to describe the radiation observed by inertial
observers at ${\cal I}^+$. The conformal transformation
$\tilde g_{\mu\nu}=\omega^2 \hat g_{\mu\nu}$ gives for the inertial radiation field
\begin{eqnarray}
    \Psi:&=&-\frac{1}{2} \lim_{\Omega \rightarrow 0}\frac{1}{\Omega}
    \tilde n^\mu {\cal Q}^\nu \tilde n^\rho {\cal Q}^\sigma 
         \tilde C_{\mu\nu\rho\sigma}   \nonumber \\
   &=&-\frac{1}{2} \omega^{-3}e^{-2i\delta} \lim_{\ell \rightarrow 0}
         \frac{1}{\ell}
    \hat n^\mu F^\nu \hat n^\rho F^\sigma \hat C_{\mu\nu\rho\sigma},
\end{eqnarray}
where ${\cal Q}^{\beta}=e^{-i\delta}\omega^{-1}F^\beta+\lambda {\tilde
n}^{\beta}$ is the same  inertial polarization dyad used in describing the news
(\ref{eq:news}).  From (\ref{eq:psisigma}), we then have
\begin{equation}
      \Psi=\frac{1}{2} \omega^{-3}e^{-2i\delta}
    \hat n^\mu F^\nu F^\rho \bigg( 
        \hat \nabla_\mu  \hat \Sigma_{\nu\rho}
    -\hat \nabla_\nu \hat \Sigma_{\mu\rho}\bigg)|_{\cal I^+} 
\label{eq:psilim}.
\end{equation}

We next need to express $\Psi$ in terms of the computational variables.
The straightforward way is to expand (\ref{eq:psilim}) as 
\begin{equation}
      \Psi=\frac{1}{2} \omega^{-3}e^{-2i\delta}
    \hat n^\mu F^A F^B \bigg( 
        \partial_\mu  \hat \Sigma_{AB}
       -\partial_A \hat \Sigma_{\mu B}
       - \hat \Gamma^\alpha_{\mu B}\hat \Sigma_{A \alpha}
       +\hat \Gamma^\alpha_{A B}\hat \Sigma_{\mu\alpha}
                      \bigg)|_{\cal I^+} 
\label{eq:psia}.
\end{equation}
and calculate the individual components of $\hat \Sigma_{\mu\nu}$ in terms of
those variables. This involves lengthy algebra, which is simplified by the
following intermediate results which hold at ${\cal I}^+$:
\begin{equation}
   \hat \Sigma_{\ell\ell} =-2\partial_\ell^2 \beta 
\end{equation}
\begin{equation}
   \hat \Sigma_{\ell A} =\frac{1}{2} e^{-2H}\partial_\ell (
     h_{AB} \partial_\ell U^B)
\end{equation}
\begin{equation}
   \hat \Sigma_{\ell u} =\frac{1}{4} e^{2H}{\cal R}
  +\frac{1}{4} D_A D^A e^{2H}- L^A \hat \Sigma_{\ell A}
\end{equation}
\begin{equation}
 (\hat \nabla^\mu \ell) (\hat \nabla^\nu \ell) \hat \Sigma_{\mu\nu} =
    \frac{1}{2} e^{-2H}(\partial_u +L^A \partial_A)(e^{-2H} D_A L^A )
\end{equation}
\begin{equation}
 (\hat \nabla^\mu \ell) \hat \Sigma_{\mu A} =
    \frac{1}{2} \partial_A (e^{-2H} D_B L^B ) 
\end{equation}
\begin{equation}
  \hat \Sigma_{AB} = \frac{1}{2}e^{-2H} (\partial_u +{\cal L}_L )c_{AB}
           +e^{-2H} D_A D_B e^{2H} 
           -\frac{1}{4}H_{AB}({\cal R} +3e^{-2H} D^C D_C e^{2H}).
\end{equation}
We use a Maple script to convert these expressions in terms of $\eth$
operators acting on the spin-weighted computational fields and construct
the final Fortran expression for $\Psi$.

In inertial Bondi coordinates, (\ref{eq:psia}) reduces to
\begin{equation}
         \Psi = \frac {1}{4} Q^A Q^B\partial_u^2  c_{AB} = \partial_u^2
               \partial_l J|_{{\cal I}}^+ . 
\end{equation}
This is related to the expression for the news function in
inertial Bondi coordinates by
\begin{equation}
       \Psi =\partial_u N.
\label{eq:PsiNu}
\end{equation}
However, as in the case of the news, the full expression (\ref{eq:psia}) for
$\Psi$ must be used in the code. This introduces additional challenges to
numerical accuracy due to the large number of terms and the appearance of third
angular derivatives.

\subsection{Linearized expressions}

The general nonlinear representation of $\Psi$ in (\ref{eq:psia}) in terms of
the computational variables is quite long but reduces to a simpler form in the
linearized approximation, i.e. to first order in perturbations off a Minkowski
background. In terms of the spin-weighted fields $J=h_{AB}q^A Q^B /2$
and $L=L^A q_A$, we find 
\begin{equation}
\Psi=-\frac{1}{2}\hat\Sigma_{\ell u}\eth L +\partial_u \hat\Sigma_J
-\frac{1}{2}\eth \hat\Sigma_u
-\frac{1}{2} \partial_u J\left(\hat\Sigma_{\ell u}+\hat\Sigma_{K}\right)
\label{e-Psil1}
\end{equation}
(evaluated at ${\cal I}^+$), where the only nonvanishing zeroth order parts of
$\hat\Sigma_{\mu\nu}$ are
\begin{equation}
\hat\Sigma_K \equiv \frac{1}{2} q^A \bar{q}^B \hat\Sigma_{AB}
=-\frac{1}{2},\;\;\hat\Sigma_{\ell u}=\frac{1}{2}
\end{equation}
and the required first order components are
\begin{equation}
\hat\Sigma_J \equiv \frac{1}{2} q^A q^B \hat\Sigma_{AB}
=\eth^2 H -\frac{1}{2}J 
+\frac{1}{2} \partial_u \partial_\ell J
\end{equation}
\begin{equation}
\hat\Sigma_u \equiv q^A \hat\Sigma_{uA}
=\frac{1}{4}\eth^2 \bar{L} +\frac{1}{4}\eth\bar{\eth}L +\frac{1}{2} L.
\end{equation}
Then (\ref{e-Psil1}) reduces to
\begin{equation}
      \Psi=\frac{1}{2}\partial_u^2 \partial_\ell J -\frac{1}{2}\partial_u J
      -\frac{1}{2}\eth L -\frac{1}{8} \eth^2( \eth \bar L +\bar \eth L)
       + \partial_u \eth^2 H.
\label{eq:linPsi}
\end{equation}

In the same approximation, the news function is given by
\begin{equation}
   N=\frac{1}{4}q^A q^B \bigg (\partial_u c_{AB} 
     +2 D_A D_B ( \omega +2H) \bigg ) =\frac{1}{2} \partial_u \partial_\ell J
      +\frac{1}{2} \eth^2(\omega +2H).
\label{eq:linN}
\end{equation}
Using the asymptotic relations
\begin{eqnarray}
     \partial_u J&=&- \eth L 
     \label{eq:linJu} \\
   \partial_u \omega &=&-\frac{1}{4}(\eth \bar L + \bar \eth L) , 
     \label{eq:linomegau}
   \end{eqnarray}
which arise from the linearized versions of (\ref{eq:asymHu}) and
(\ref{eq:omegadot}), it is easy to see that (\ref{eq:PsiNu}), i.e. $\Psi
=\partial_u N$, still holds in the linearized approximation. (In the nonlinear
case, the derivative along the generators of ${\cal I}^+$ is $\hat n^\mu
\partial_\mu =e^{-2H}(\partial_u +L^A \partial_A)$ and (\ref{eq:PsiNu}) must be
modified accordingly.)

The linearized expressions (\ref{eq:linPsi}) and (\ref{eq:linN}) provide a
starting point to compare the advantages between computing the radiation via the
Weyl component $\Psi$ or the news function $N$. The troublesome  terms
involve $L$, $H$ and $\omega$, which all vanish in inertial Bondi coordinates.
One main  difference is that $\Psi$ contains third order angular
derivatives, e.g. $\eth^3 \bar L$, as opposed to second angular derivatives for
$N$. This means that the smoothness of the numerical error is more crucial in
the $\Psi$ approach. Balancing this, another main difference is that $N$
contains the $\eth^2 \omega$ term, which is a potential source of numerical
error since $\omega$ must be propagated across patch boundaries via
(\ref{eq:omegadot}).

\subsection{Summary of the gravitational radiation calculation}
\label{sec:sum}

The characteristic Einstein equations are evolved in a domain between an inner
radial boundary at the interior worldtube, and an outer boundary at future null
infinity. Initial data for $J(u,r,x^A)$ is required at $u=0$. This data is
constraint-free so that, in the absence of an exact solution or other
prescription of the data, we can simply set $J(0,r,x^A)=0$. Alternatively, in
order to reduce spurious initial radiation, we can set the Newman-Penrose Weyl
tensor component $\Psi_0(0,r,x^A) =0$, which determines $J(0,r,x^A)=0$ when
continuity conditions are imposed at the inner worldtube. The metric data from a
Cauchy evolution is interpolated onto the inner worldtube to extract the
boundary data for the characteristic evolution. This extraction process involves
carrying out the complicated Jacobian transformation between the Cartesian
coordinates used in the Cauchy evolution and the spherical null coordinates 
used in the characteristic evolution. The full details are given in~\cite{ccm}.
The result is boundary data for $J,\beta,U,Q,V$ on the worldtube, which supply
the integration constants for a radial numerical integration of (\ref{eq:beta}),
(\ref{eq:wua}), (\ref{eq:wq}) and (\ref{eq:ww}), in that order. Given the
initial data $J(0,r,x^A)$, this leads to complete knowledge of the metric on the
initial null cone. Then (\ref{eq:wev}) gives an expression for $J_{,ur}$, which
is used to determine $J$ on the ``next'' null cone, so that the process can be
repeated to yield the complete metric throughout the domain, which extends to
${\cal I}^+$.

Before the gravitational radiation is calculated from the metric in the
neighborhood of ${\cal I}^+$, it is necessary to compute the auxiliary
variables $\omega(u,x^A)$ and $\delta(u,x^A)$ which determine the inertial
polarization dyad in which to measure the news function $N$ or Weyl component
$\Psi$. Given a solution of (\ref{eq:conf}) for the initial value of
$\omega(0,x^A)$, its evolution is computed by integrating (\ref{eq:omegadot}).
(If $J=0$ initially, then $\omega=1$ is  the solution to (\ref{eq:conf}).
Otherwise, $\omega$ is initiated by solving a 2-dimensional elliptic equation.)
Similarly, fixing the initial polarization basis by  $\delta(0,x^A) =0$, its
evolution is computed by integrating (\ref{eq:evphase}). Then the news $N$ is
given by (\ref{eq:news}) (or in spin-weighted form by the formulas in Appendix
B of Ref.~\cite{highp}) and $\Psi$ is given by (\ref{eq:psia}).

The above procedure computes $N$ or $\Psi$ as functions of the code coordinates
$(u,x^A)$, rather than inertial coordinates. In the linearized case, which is
used for the tests in Sec.~\ref{sec:extraction}, the change to inertial
coordinates is a second-order effect that can be neglected. However, that is not
the case in general and the full procedure is described in Sec. IV B of
Ref.~\cite{highp}.

\section{Patching the sphere}
\label{sec:patches}

The nonsingular description of smooth tensor fields on the sphere requires more
than a single coordinate patch. Here we consider the stereographic treatment
which uses 2 coordinate patches, and the cubed-sphere treatment, which uses 6
patches. In both cases the metric $q_{AB}$ of the unit sphere is expressed in
terms of a complex dyad $q_A$ (satisfying $q^Aq_A=0$, $q^A\bar q_A=2$,
$q^A=q^{AB}q_B$, with $q^{AB}q_{BC}=\delta^A_C$ and $ q_{AB}
=\frac{1}{2}\left(q_A \bar q_B+\bar q_A q_B\right)$). The dyads for each patch
are related by spin transformations at points common
to more than one patch. 

\subsection{Circular stereographic patches}

In stereographic coordinates, the sphere is covered with 
two patches, one for each hemisphere. 
The North hemisphere is covered by the complex stereographic coordinate 
$\xi_N=\eta_N+i\rho_N$, which is related to
standard $(\theta, \phi)$ angular coordinates by
$\xi_N=\tan(\theta/2)e^{i\phi}$ and which
is regular on the entire sphere except for the South pole. 
The South hemisphere is covered by the complex stereographic coordinate
$\xi_S=1/\xi_N=\eta_S+i\rho_S=\cot(\theta/2)e^{-i\phi}$, 
which is singular at the North pole. 
Every point on the sphere is covered by at least one of the patches, and there 
is a region around the equator where points are covered by both patches. 
In this overlap region between the two patches, a scalar function $F$ with 
value $F_N(\xi_N)$ on the North patch has the value 
$F_S(\xi_S=1/\xi_N)$ on the South patch. For a function $F$ of spin-weight 
$s$, $F_S(\xi_S=1/\xi_N)=F_N(\xi_N)(-1)^se^{-2is\Phi}$.

In the $x^A=(\eta,\rho)$ coordinates, the unit sphere
metric in each patch is given by
\begin{equation}
q_{AB} dx^A dx^B = \frac{4}{P^2}(d\eta^2+d\rho^2),
\end{equation}
where
\begin{equation}
        P=1+\eta^2+\rho^2.
\end{equation}
The equator corresponds to the circle
\begin{equation}
\sqrt{\eta^2+\rho^2} = 1.
\end{equation}  We fix
the dyad by the explicit choice
\begin{equation}
      q^A=\frac{P}{2}(1,i)\, , \quad i=\sqrt{-1}.
\end{equation}

In the composite mesh method, all boundary points of one patch are interior
points of another patch. The overlapping of the patches is  key to the
stability of this method. The two stereographic
coordinate patches must both extend
beyond the equator. In the scheme originally used to implement the computational
$\eth$-formalism~\cite{eth} in the characteristic code, the North and South
patches were represented by square $(\eta,\rho)$ grids. In the scheme implemented for
meteorological studies~\cite{browning}, circular masks are applied so that the
computational grids extend only a few zones beyond the equator. Here we adopt
this circular grid boundary but place it a fixed geometrical distance past the
equator, i.e. the grid boundary for the North patch is a circle lying in the
South patch. The finite buffer zone between the equator and the grid boundary
allows for angular dissipation, as developed in Sec.~\ref{sec:angdiss}, to damp
the high frequency intergrid interpolation error before it crosses the equator.
This protects measurements of the news function (or $\Psi$) in the North patch,
which involve two (or three) angular derivatives, from substantial
contamination by the interpolation error at the patch boundary. 

We discretize the stereographic coordinates according to
\begin{equation}
\eta_i = -1 + (i-O-1)\Delta \, , \quad
\rho_j = -1 + (j-O-1)\Delta
\label{eq:sqgrid}
\end{equation}
where, following the notation in~\cite{browning}, $O$ is the number of points
(overlapping points) by which each grid extends beyond the equator 
and the indices range over  $1 \le i,j \le M+1+2O$, with $M^2$ being the number of
grid points inside the equator. The grid spacing  $\Delta$ depends on $M$ according to 
\begin{equation}
\Delta = \frac{2}{M}.
\end{equation}
The square grid determined by (\ref{eq:sqgrid}) ranges over
\begin{equation}
       (\eta_i,\rho_j) \in \left( -1-O\Delta, 1+O\Delta \right) 
\label{eq:qpgrid}
\end{equation}
in each patch.

In the original square patch method~\cite{eth}, the evolution algorithm is
applied to the entire set of points in the square $(\eta,\rho)$ grid, with the
field values at the resulting ghost points supplied by interpolation from the
other patch. In the circular patch method~\cite{browning}, the evolution
algorithm is only applied to points inside a circle $r=\sqrt{\eta^2+\rho^2}$,
where $r>1$ so that the boundary lies a small distance past the equator. In
convergence tests, the number of overlap points determined by $O$ is adjusted so
that $r$ is at a fixed position for all grids, i.e. $O$ scales as $1/\Delta$.
The grid points outside this circle are either ghost points or inactive. The
circular patch is clearly more economical than using a  square patch and avoids
the error introduced by the large stereographic grid stretching near the corners
of the square.

When the finite-difference stencil is used near the boundary of the active grid
points, field values required at the ghost points outside a circular patch are
interpolated from values at interior  points of the opposite patch. The
algorithm for determining the value of a scalar function $F_N$ at a ghost point
in the North patch starts with the determination of the ghost point's 
coordinates in the overlapping South patch, followed by the interpolation of the
value of the function $F_S$ at the ghost point, i.e. the $F_N$ ghost point
values are obtained by interpolation via the $F_S$ active grid values. 

Let $R_E$ be the width of the finite-difference stencil divided by $2 \Delta$.
In the circular patch method, we define the {\em active} finite difference
grid, i.e. the grid points to which the
evolution algorithm is applied, by
\begin{equation}
\sqrt{\eta_i^2+\rho_j^2} \le 1 + (O-R_E)\Delta,
\label{eq:evmask}
\end{equation} 
where $O > R_E$. Stability of the composite mesh method requires that the
interpolation stencil for the ghost points for one patch lies below the equator
in the other patch. Those requirements give a  minimum value of $O$ but a larger
value may be necessary to establish an effective buffer zone for the dissipation
to attenuate the interpolation error before it enters the opposite patch. The 
optimum value of $O$ in order to avoid instability or inaccuracy, needs to be
established by experiment. (Too large a value would lead to inaccuracy due to
the stretching of the stereographic grid.)

\subsection {The cubed sphere}

In the cubed-sphere approach, developed for meteorological studies
in~\cite{ronchi} and later for numerical relativity in~\cite{Thornburgah}, 6
coordinate patches on the sphere are obtained by projecting the 6 faces of a
circumscribed cube. The method has recently been applied to characteristic
evolution in~\cite{reisswig} and independently in~\cite{roberto}. Here we follow
the notation of~\cite{reisswig}, except we denote the angular coordinates by
$(\phi_1,\phi_2)$ (rather than by $(\rho,\sigma)$). In addition, in order to
ensure that the coordinates and dyads on each patch are consistently
right-handed, with the vector cross-product vector pointing out of the sphere,
we introduce some sign changes in the conventions used in~\cite{reisswig} for
the coordinate transformations between the patches and in the definition of the
dyad $q^A$. These conventions simplify the interpatch transformation of
spin-weighted quantities.

Given Cartesian coordinates $(x,y,z)$, we define
angular coordinates $x^A=(\phi_1,\phi_2)$ on the 2-sphere $x^2+y^2+z^2=1$
by means of the six patches $(x_\pm,y_\pm,z_\pm)$, where
\begin{eqnarray}
x_\pm&:&\;
\phi_1=\arctan \left(\pm \frac{y}{x}\right),
\phi_2=\arctan \left( \frac{z}{x}\right) 
\nonumber \\
y_\pm&:&\;
\phi_1=\arctan \left( \pm\frac{z}{y}\right),
\phi_2=\arctan \left( \frac{x}{y}\right)  \nonumber \\
z_\pm&:&\;
\phi_1=\arctan \left( \pm \frac{x}{z}\right),
\phi_2=\arctan \left( \frac{y}{z}\right).
\label{e-six2C}
\end{eqnarray}

In each patch, the range of the coordinates is
$-\pi/4 \le\phi_1,\phi_2\le\pi/4$ and the metric is
\begin{equation}
ds^2=\left(1- \sin^2 \phi_1 \; \sin^2 \phi_2 \right)^{-2} 
\bigg(
\cos^2\phi_2  \; d\phi_1^2 + \cos^2\phi_1 \;  d\phi_2^2
- \frac{1}{2}  \sin(2\phi_1) \sin(2\phi_2) \; d\phi_1 \, d\phi_2
\bigg).
\label{e-m1}
\end{equation}
As a simple dyad representing (\ref{e-m1}), we choose 
\begin{equation}
q_A=\bigg(
\frac{(\theta_c-i \theta_s)\cos\phi_2}{4\theta_c^2 \theta_s^2},
\frac{(\theta_c+i \theta_s)\cos\phi_1}{4\theta_c^2 \theta_s^2}
\bigg),\;\;
q^A=\bigg(
2\theta_c\theta_s\frac{\theta_s
-i \theta_c}{\cos\phi_2},
2\theta_c\theta_s\frac{\theta_s
+i \theta_c}{\cos\phi_1}
\bigg),
\label{e-dyad}
\end{equation}
where
\begin{equation}
\theta_c=\sqrt{\frac{1-\sin\phi_1\sin\phi_2}{2}},
\theta_s=\sqrt{\frac{1+\sin\phi_1\sin\phi_2}{2}}.
\end{equation}
The operator $\eth$ acting on
a field $S$ with spin-weight $s$ is
$\eth S=q^A \partial_A S + s \Gamma S$ where, with the present conventions,
\begin{eqnarray}
\Gamma&=&
\frac{\cos^2\phi_1\cos^2\phi_2(\sin\phi_1+\sin\phi_2)
     +(\cos^2\phi_1-\cos^2\phi_2)(\sin\phi_2-\sin\phi_1)}
             {4 \theta_c \cos\phi_2 \cos\phi_1} \nonumber \\
     &-&i\frac{\cos^2\phi_1 \cos^2\phi_2 (\sin\phi_1-\sin\phi_2)
     +(\cos^2\phi_2-\cos^2\phi_1)(\sin\phi_1+\sin\phi_2)}
            {4 \theta_s \cos\phi_2 \cos\phi_1}.
\label{e-Gamma}
\end{eqnarray}

We introduce ghost zones in the usual manner along the boundaries of each patch,
and couple the patches together by interpolating the field variables from
neighboring patches to each ghost point. With the definition (\ref{e-six2C}),
the angular coordinate~$\phi_1$ or~$\phi_2$ perpendicular to an interpatch
boundary is always common to both adjacent patches. This greatly simplifies
interpatch interpolation, since it only needs to be done in
1~dimension, parallel to the boundary.

\section{Comparison between stereographic and cubed-sphere methods}
\label{sec:stests}

We carry out a test of wave propagation on the sphere to compare the accuracy of
using circular stereographic patches with the cubed-sphere methods, with
emphasis on the accuracy of the angular derivatives required in waveform
extraction by the news and $\Psi_4$ approaches. The test allows direct
comparison between the stereographic and cubed-sphere treatments without
introducing the complications of characteristic evolution and the explicit
calculations of the news or $\Psi$.

The test is based upon solutions to the 2D wave equation
\begin{equation}
    - \partial_t ^2 \Phi + \eth \bar \eth \Phi = 0,
\end{equation}
where $\Phi =cos(\omega t) Y_{\ell m}$, $\omega =\sqrt{\ell(\ell+1)}$
and $Y_{\ell m}$ are spherical harmonics.

For the case $\ell=m=2$, we compare test results for the stereographic
grid with circular patches and the
cubed-sphere grid. For the stereographic grid, the simulations are run with
$M^2$ grid points in each patch, for $M=100$ and $M=120$. The
corresponding cubed sphere runs keep the number of  grid-cells covering the
sphere the same as for the stereographic case, not counting those cells that
overlap with another patch. For $M^2$ stereographic grid points there are
$\approx \pi M^2 / 4$ grid-cells inside the equator on each hemisphere. In the
cubed sphere grid, with $N^2$ points per patch, the entire sphere is covered by
$6 \times N^2$ points. Equating the number of cells for the entire sphere gives
$N^2 \approx (\pi/12) M^2$.  The above values of $M$ then
correspond to $N = 51, 61$. 
The tests are run until $t=120$.  

Angular dissipation is necessary for the stability of the stereographic runs. 
For grid size $\Delta$, it was added in the finite-difference form 
\begin{equation}
    \partial_t^2 \Phi \rightarrow  \partial_t^2 \Phi 
         + \epsilon \Delta^3 {\cal D}^4 \partial_t\Phi , 
\end{equation}
where ${\cal D}^4 \Phi =\left (\frac{P^2}{4}({\cal D}_{+\eta}{\cal D}_{-\eta}
+{\cal D}_{+\rho}{\cal D}_{-\rho})\right )^2\Phi $, where ${\cal D}_{+}$ (or
${\cal D}_{-}$) indicates the forward (or backward) difference operator in the
indicated direction. Experimentation with tuning the dissipation revealed that a
small value $\epsilon = 0.01$ is sufficient to suppress high  frequency error.
The finite difference stencil (taking dissipation into account) has width
$R_E=2$. Using a $4th$ order Lagrange interpolator and by tuning the number $O$
of overlapping points, we obtained good results with $O=5$. Angular dissipation
was not used in the cubed sphere runs.

We use the $L_\infty$ norm to measure the error 
\begin{equation}
       {\cal E}(\Phi) =||\Phi_{numeric}-\Phi_{analytic}||_\infty.
\end{equation}
We measure the convergence rate for ${\cal E}(\Phi)$ at a given time $t$,
for two grid sizes $\Delta_1$ and  $\Delta_2$, by
\begin{equation}
    {\cal R} = \frac{\log_2 \big ({\cal E}(\Phi)_{\Delta_1}
                            / {\cal E}(\Phi)_{\Delta_2} \big )} 
                {\log_2 \big (\Delta_2 / \Delta_1 \big )}.
\end{equation}
Convergence rates for other quantities are measured analogously. For a given
grid, we measure the error for the circular patches in the North
hemisphere; while for the cubed-sphere method we measure the error on the
$(+x,+y,+z)$ patch, excluding ghost points at the edges of the patch. The
finite difference approximations for the codes are designed to be second
order accurate.

\begin{figure}[htp] %  figure placement: here, top, bottom, or page
   \centering
   \psfrag{time}{t}
   \psfrag{circular}{circular patch}
   \psfrag{sixpatch}{cubed-sphere}
   \psfrag{error}[c][c]{${\cal E}(\Phi)$ }
   \includegraphics*[width=10cm]{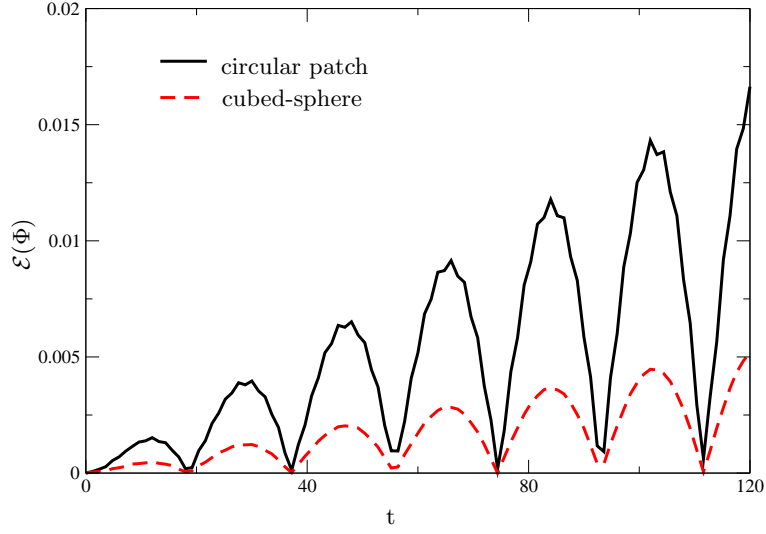} 
   \caption{Comparison of the $L_\infty$ error ${\cal E}(\Phi)$ vs $t$, 
	    for the highest resolution runs using the circular patch method
            and the cubed-sphere method. }
   \label{fig:Phin_error}
\end{figure}

Excellent second-order convergence of ${\cal E}(\Phi)$, based upon the
 $M=100$ and $M=120$ grids, is evident for both methods from the results listed in
Table~\ref{tab:ConvPhi}. The time dependence of the error plots in
Fig.~\ref{fig:Phin_error}, shows that the cubed-sphere error is
$\approx \frac{1}{3}$ the stereographic error. 

\begin{table}[htp]
  \begin{center}
    \begin{tabular}[c]{|c|c|c|c|c|}
\hline
  ALGORITHM &
  t=1.2 &
  t=12 &
  t = 102 &
  t = 120

    \\ \hline \hline
%%%%%
circular patch        & $2.002$ & $1.988$ & $1.994$ & $1.999$
\\ \hline
%%%%%
%%%%%
cubed-sphere          & $1.994$ & $1.970$ & $1.982$ & $1.985$
\\ \hline
%%%%%
\end{tabular}
    \caption{Convergence rates for ${\cal E}(\Phi)$, obtained with 
             the $L_\infty$ norm using the $M=100$ and $M=120$ grids.}
    \label{tab:ConvPhi}
  \end{center}
\end{table}

\begin{figure}[htp] %  figure placement: here, top, bottom, or page
   \centering
   \psfrag{time}{t}
   \psfrag{circular}{circular patch}
   \psfrag{sixpatch}{cubed-sphere}
   \psfrag{error}[c][c]{${\cal E}(\eth^2 \Phi)$ }
   \includegraphics*[width=10cm]{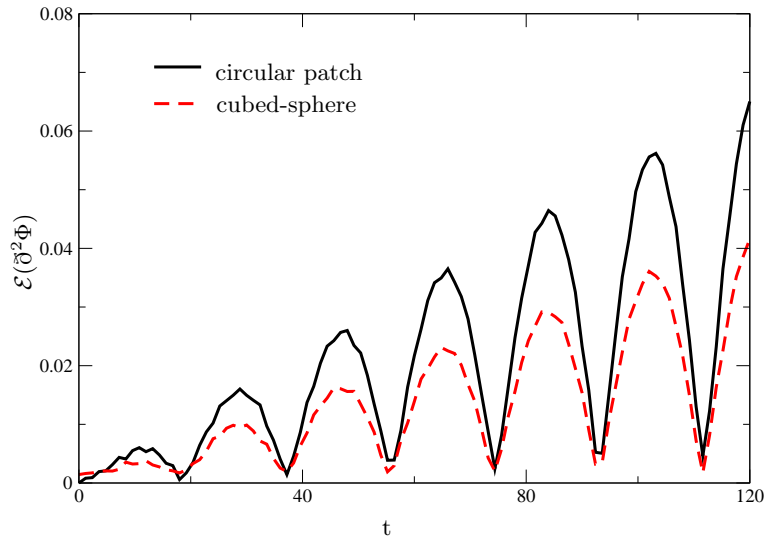} 
   \caption{The $L_\infty$ error ${\cal E}(\eth^2 \Phi)$ vs
	    $t$ for the highest resolution runs is compared using
            the circular patch method and the cubed-sphere method. }
   \label{fig:eth2Phin_error}
\end{figure}

A more important test to assess the error relevant to gravitational wave
extraction is to measure the error in  $\eth^2\Phi$, since second angular
derivatives enter in the  computation of the Bondi news.  The convergence rates,
measured with the $L_\infty$ norm, are shown in Table~\ref{tab:Conveth2Phi}. The
circular patch and cubed-sphere results indicate clean second order convergence
up to the final run time at $t=120$. The plots of the error versus time in
Fig.~\ref{fig:eth2Phin_error} show that the error for the cubed-sphere is about
$\frac{2}{3}$rd the stereographic error. 

\begin{table}[htp]
  \begin{center}
    \begin{tabular}[c]{|c|c|c|c|c|}
\hline
  ALGORITHM &
  t=1.2 &
  t=12 &
  t = 102 &
  t = 120 

    \\ \hline \hline
%%%%%
circular patch        & $2.022$ & $1.945$ & $1.992$ & $2.006$
\\ \hline
%%%%%
%%%%%
cubed-sphere          & $1.954$ & $2.019$ & $1.997$ & $1.971$
\\ \hline
%%%%%
\end{tabular}
    \caption{Convergence rates for the $L_\infty$ error ${\cal E}(\eth^2\Phi)$
     for various times $t$, obtained using the two highest resolution runs.
          }
    \label{tab:Conveth2Phi}
  \end{center}
\end{table}

Similarly, it is important for the purpose of gravitational wave extraction
using the Weyl tensor to measure the error in $\bar \eth \eth^2\Phi$, since third angular
derivatives enter into the
computation of $\Psi$. The corresponding convergence rates are shown in
Table~\ref{tab:Convethbeth2Phi}. Now the cubed-sphere method shows poor
convergence  of the $L_\infty$ error at early times. The underlying error is
generated at the corners where three patches meet, as indicated by the improved
convergence rate measured using the $L_2$ norm. Apparently some built-in dissipation of the
evolution algorithm smooths this patch-boundary error and second order
convergence is evident by  by $t=102$. To a much smaller extent,
the $L_\infty$ error for the circular
patch method also shows some deviation from second order convergence at early
times, but clean second order convergence is evident by
$t=12$.

\begin{figure}[htp] %  figure placement: here, top, bottom, or page
   \centering
   \psfrag{time}{t}
   \psfrag{circular}{circular patch}
   \psfrag{sixpatch}{cubed-sphere}
   \psfrag{error}[c][c]{${\cal E}(\bar \eth \eth^2 \Phi)$ }
   \includegraphics*[width=10cm]{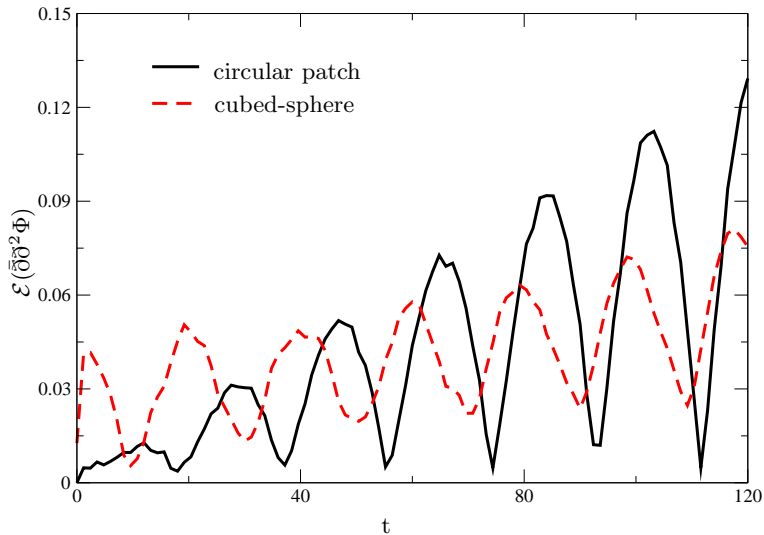} 
   \caption{The error ${\cal E}(\bar \eth \eth^2
	    \Phi)$ vs $t$ is compared for the highest resolution runs using 
            the circular patch method and the cubed-sphere method. 
            }
   \label{fig:ethbeth2Phin_error}
\end{figure}

The magnitude of the $L_\infty$ error in $\bar \eth \eth^2\Phi$ is plotted vs
time in Fig.~\ref{fig:ethbeth2Phin_error}. Until about $t=60$,
the cubed-sphere method has the largest error. But at the end of the run
at $t=120$ the cubed-sphere error is about $\frac{4}{5}$ the
stereographic error.

\begin{figure}[htp] %  figure placement: here, top, bottom, or page
    \centering
    \psfrag{xmtickgz}[c][b]{$-\pi/4$}
    \psfrag{xztickgz}[c][b]{$0$}
    \psfrag{xptickgz}[c][l]{$+\pi/4$}
    \psfrag{ymtickgz}[c][r]{$-\pi/4$}
    \psfrag{yztickgz}[c][r]{$0$}
    \psfrag{yptickgz}[c][r]{$+\pi/4$}

    \psfrag{xmtick}[c][b]{$-1$}
    \psfrag{xztick}[c][b]{$0$}
    \psfrag{xptick}[c][b]{$+1$}
    \psfrag{ymtick}[c][b]{$-1$}
    \psfrag{yztick}[c][b]{$0$}
    \psfrag{yptick}[c][b]{$+1$}

    \includegraphics*[width=8.5cm]{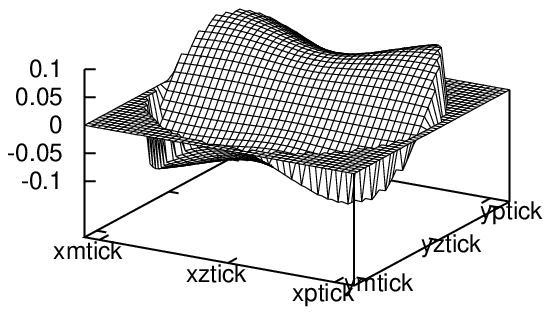} 
    \includegraphics*[width=8.5cm]{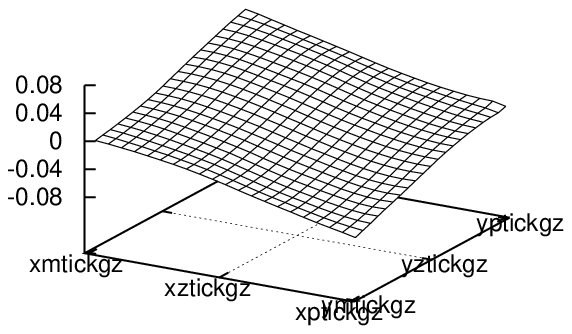} 
    \caption{2D snapshots of the error in 
             $\bar \eth \eth^2\Phi$ at $t=120$ for the 
             on the North hemisphere for the circular patch method (left plot),
             and for a cubed-sphere patch (right plot). 
	     For the sake of plot clarity these
	     2D snapshots use only every third data-point
	     along each axis. The third angular derivatives are smooth for
	     both methods
             }
    \label{fig:ethbeth2Phin_2Derror}
 \end{figure}

Surface plots of the error at the final run time are shown in
Fig.~\ref{fig:ethbeth2Phin_2Derror}. The circular patch and cubed-sphere errors
are both smooth, as would be expected of the second order truncation error
arising from the finite differencing. For the circular patch, this shows that
dissipation in the buffer zone surrounding the equator effectively guards
against the high frequency error introduced at the patch boundary. Our results
for the stereographic method justify its use in the comparison of the news and
Weyl tensor extraction strategies in Sec.~\ref{sec:extraction}.

\begin{table}[htp]
  \begin{center}
    \begin{tabular}[c]{|c|c|c|c|c|}
\hline
  ALGORITHM &
  t=1.2 &
  t=12 &
  t = 102 &
  t = 120 

    \\ \hline \hline
%%%%%
circular patch, $L_\infty$ norm       & $2.278$ & $2.032$ & $1.988$ & $2.009$
\\ \hline
%%%%%
%%%%%
cubed-sphere, $L_\infty$ norm & $1.108$ & $0.882$ & $2.009$ & $1.959$
\\ \hline
cubed-sphere, $L_2$ norm & $1.883$ & $1.983$ & $1.981$ & $1.959$
\\ \hline

%%%%%
\end{tabular}
    \caption{Convergence rates for the error ${\cal E}(\bar \eth
	     \eth^2\Phi)$. For the cubed-sphere method the dominant error arises
	     at the patch corners, which is revealed by the comparison of the
             $L_\infty$ and $L_2$ errors. This can be understood in terms of the
             the inter-patch interpolation stencil which is partially
             off-centered near the corners, where the error is greatest.
             The inherent numerical dissipation of the evolution algorithm
             keeps this localized, non-smooth noise from growing,
             while smoother error from other regions grows linearly in time
             (see Fig.~\ref{fig:ethbeth2Phin_error}). The net
             effect is that at late times both the $L_\infty$  and the
             $L_2$ norms of the error in the third derivative
             show second order convergence, while early in the evolution
             the $L_\infty$ shows only first order convergence.
          }
    \label{tab:Convethbeth2Phi}
  \end{center}
\end{table}

\clearpage

\section{Comparison of news and Weyl tensor extraction}
\label{sec:extraction}

Here we  compare the accuracy of waveform  extraction by computing the news
function $N$ or the Weyl tensor component $\Psi$ in a linearized gravitational
wave test problem. The computations are carried out by the procedure described
in Sec.~\ref{sec:sum}. In accord with (\ref{eq:PsiNu}), the $\Psi$ computation
yields an alternative numerical value for the news
\begin{equation}
     N_\Psi = N|_{u=0} +\int_0^u \Psi du,
\label{eq:npsi}
\end{equation}
where $N=N_\Psi$ in the analytic problem.
We compare the two extraction methods in terms of the errors
in $N$ and $N_\Psi$ obtained using the stereographic method.

We base the test on a class of solutions in Bondi-Sachs form to the linearized
vacuum Einstein equation on a Minkowski background given in Sec. 4.3
of~\cite{BS-lin}. The solution allows us to make convergence checks of the
Bondi-Sachs metric quantities as well as the news function. The solutions are
expressed in terms of spin-weighted spherical harmonics ${}_s Y_{\ell
m}$~\cite{newp,golm}, modified to avoid mixing of the $m$ and $-m$ components
when extracting the real part according to~\cite{mod}   
\begin{equation}
{}_s R_{\ell m} = \frac{1}{\sqrt{2}} \left[{}_s Y_{\ell m}
   +(-1)^m {}_s Y_{\ell -m}\right] \mbox{ for } m>0,\;
{}_s R_{\ell m} = \frac{i}{\sqrt{2}} \left[(-1)^m{}_s Y_{\ell m} 
   -{}_s Y_{\ell -m} \right]\mbox{ for }  m<0 ,\;
{}_s R_{\ell 0} = {}_s Y_{\ell 0}.
\end{equation}
Ref.~\cite{mod} gives explicit expressions for the ${}_s R_{\ell m}$
in stereographic coordinates.

Following~\cite{reisswig}, these
linearized  solutions have Bondi-Sachs variables
\begin{eqnarray}
J&=& \sqrt{(\ell -1)\ell(\ell+1)(\ell+2)}\;
{}_2 R_{\ell m} \Re(J_\ell(r) e^{i\nu u}), \;
U= \sqrt{\ell(\ell+1)}\;{}_1R_{\ell m} \Re(U_\ell(r) e^{i\nu u}),
\nonumber \\
\beta&=& R_{\ell m} \Re(\beta_\ell e^{i\nu u}),
\; W_c= R_{\ell m} \Re(W_{c\ell}(r) e^{i\nu u}),
\label{e-an}
\end{eqnarray}
where $W_c$ determines the perturbation in $V$.
Here $J_\ell(r)$, $U_\ell(r)$, $\beta_\ell$, $W_{c\ell}(r)$ are in general
complex, and taking the real part leads to $\cos(\nu u)$ and $\sin(\nu u)$
terms. The quantities $\beta$ and $W_c$ are real, while $J$ and $U$ are
complex. We
require a solution that is well-behaved at future null infinity, and is
well-defined for $r \ge r_0>0$, where $r_0$ is the inner boundary.
We find in the case $\ell=2$
\begin{eqnarray}
\beta_2&=&\beta_0 \nonumber \\
J_2(r)&=&\frac{24\beta_0 +3 i \nu C_1 - i \nu^3 C_2}{36}+\frac{C_1}{4 r}
       -\frac{C_2}{12 r^3} \nonumber \\
U_2(r)&=&\frac{-24i\nu \beta_0 +3 \nu^2 C_1 - \nu^4 C_2}{36} +\frac{2\beta_0}{r}
       +\frac{C_1}{2 r^2} +\frac{i\nu C_2}{3 r^3} +\frac{C_2}{4 r^4} \nonumber \\
W_{c2}(r)&=&\frac{24i\nu \beta_0 -3 \nu^2 C_1 + \nu^4 C_2}{6} 
           +\frac{3i\nu C_1 -6\beta_0-i\nu^3 C_2}{3r}
           -\frac{\nu^2 C_2}{r^2} +\frac{i\nu C_2}{r^3} +\frac{C_2}{2r^4},
\label{e-NBl2}
\end{eqnarray}
with the (complex) constants $\beta_0$, $C_1$ and $C_2$ freely specifiable.
In the case $\ell=3$
\begin{eqnarray}
\beta_2&=&\beta_0 \nonumber \\
J_3(r)&=&\frac{60\beta_0 +3 i \nu C_1 + \nu^4 C_2}{180}+\frac{C_1}{10 r}
       -\frac{i \nu C_2}{6 r^3} -\frac{C_2}{4r^4}
\nonumber \\
U_3(r)&=&\frac{-60i\nu \beta_0 +3 \nu^2 C_1 - i \nu^5 C_2}{180}
         +\frac{2\beta_0}{r} +\frac{C_1}{2 r^2}
         -\frac{2\nu^2 C_2}{3 r^3} +\frac{5 i \nu C_2}{4 r^4}
         + \frac{C_2}{r^5}\nonumber \\
W_{c3}(r)&=&\frac{60 i \nu \beta_0 -3 \nu^2 C_1 + i\nu^5 C_2}{15} 
           +\frac{i\nu C_1 -2\beta_0+\nu^4 C_2}{3r}
           -\frac{i2\nu^3 C_2}{r^2} -\frac{4i\nu^2 C_2}{r^3}
           +\frac{5\nu C_2}{r^4}+\frac{3 C_2}{r^5}.
\label{e-NBl3}
\end{eqnarray}

The  news $N$ for the linearized wave is given by
\begin{equation}
N=\Re\left(  e^{i\nu u}\lim_{r \rightarrow \infty}
\left(-\frac{\ell(\ell+1)}{4}J_{\ell}-\frac{i\nu}{2} r^2 J_{\ell,r}\right)
 + e^{i\nu u}\beta_\ell \right) \sqrt{(\ell-1)\ell(\ell+1)(\ell+2)}
\;{}_2R_{\ell m},
\label{e-Nl}
\end{equation}
corresponding
For the cases $\ell=$2 and 3, this gives
\begin{equation}
\ell=2:\;\;N=\Re\left(\frac{i\nu^3 C_2}{\sqrt{24}} e^{i\nu u}\right)
  \;{}_2R_{2m};\;\;\;
\ell=3:\;\;N=\Re\left(\frac{-\nu^4 C_2}{\sqrt{30}} e^{i\nu u}\right)
  \;{}_2R_{3m}.
\label{e-N}
\end{equation}
For the linearized case $\Psi=N_{,u}$, which gives
\begin{equation}
\ell=2:\;\;\Psi=\Re\left(\frac{-\nu^4 C_2}{\sqrt{24}} e^{i\nu u}\right)
  \;{}_2R_{2m};\;\;\;
\ell=3:\;\;\Psi=\Re\left(\frac{-i\nu^5 C_2}{\sqrt{30}} e^{i\nu u}\right)
  \;{}_2R_{3m}.
\end{equation}

\subsection{Test specifications}

Tests were run with the solution parameters $\nu=1$ and $m=0$
for the cases $\ell=2$ and $\ell=3$, with
\begin{eqnarray}
   C_1&=&3\cdot 10^{-6}\, , \quad C_2 =10^{-6}\, , 
           \quad  \beta_0 =i\cdot 10^{-6} \quad (\ell=2) \\
   C_1&=&3\cdot 10^{-6}, \quad C_2 =i \cdot 10^{-6}\, , 
           \quad   \beta_0 =i\cdot 10^{-6} \quad (\ell=3). 
\end{eqnarray}
The inner worldtube boundary was placed at $r_0=2$ corresponding to
a compactified radial coordinate $x_0=r_0/(R+r_0)\approx .1888$, where
we have set the scale parameter $R=9$. 

For the convergence measurements, the $(\eta,\rho,x)$ grid consisted of $M^3$ points,
with $M=100$ and $M=120$. The boundary of the circular patches were fixed at 
$\sqrt{\eta^2+\rho^2}=1.4$. The runs were stopped at $t=100$. The
$L_\infty$ and $L_2$ error norms were computed on the North hemisphere, using
the values from the North patch.

Angular dissipation was applied only to the circular patch runs, with the
dissipation coefficients $\epsilon_x=0.009$, $\epsilon_u=0.0009$,
$\epsilon_Q=\epsilon_W=0.00001$. The weighting function ${\cal W}$ for
application of the dissipation was taken to be a unit step function which
vanishes for $\sqrt{\eta^2+\rho^2}\ge 1.3$.

We present output data for the real parts of $J$, $N$
and $N_\Psi$, For the $m=0$ case, these quantities correspond to
a pure $\oplus$ polarization mode. For comparison purposes, we include
results for the circular patch without dissipation and the original
square patch treatment.

\subsection{Test results for $J$}

\begin{figure}[htp] %  figure placement: here, top, bottom, or page
    \centering
    \psfrag{xlabel}{x}
    \psfrag{ylabel}{${\cal E}(J)$}
    \includegraphics*[width=10cm]{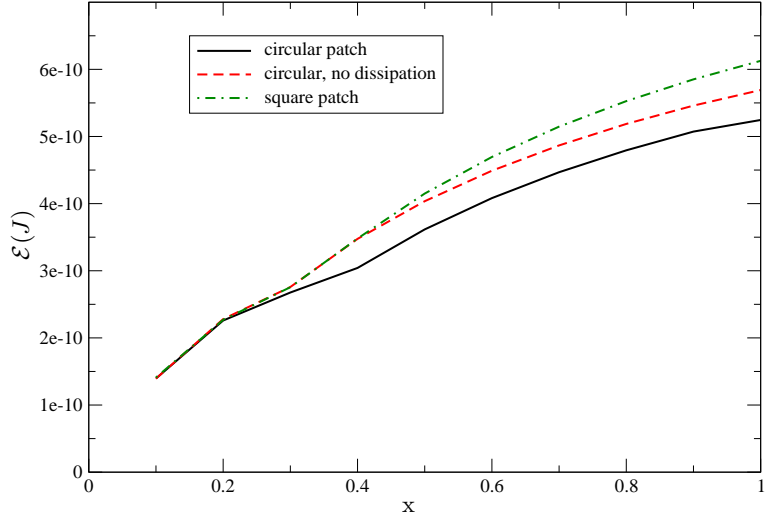} 
    \caption{The $L_\infty$ error ${\cal E}(J)$ plotted vs $x$ at $t=100$ for
             runs with the circular patch method (with and without dissipation)
             and with the square patch method.}
    \label{fig:InfErrJnorm}
 \end{figure}

We first present test results for $J$, which is a typical metric quantity
entering into the waveform calculation. Figure \ref{fig:InfErrJnorm} show the
$L_\infty$ norm over the North hemisphere of the error ${\cal E}(J)$ vs the
compactified radial coordinate $x$ at the end of the run at $t=100$ for the
$\ell=2$ wave. The figure compares runs made with the circular patch method
(dissipation applied) with runs without dissipation and runs with the original
square patch method. The plots show that angular dissipation reduces the error.
This will become more evident in the later test results for the news in which
higher angular derivatives are involved. An important feature of the plots is
that in all cases the error increases monotonically and takes it maximum value
at ${\cal I}^+$ ($x=1$), as would be expected of the radial marching algorithm.
This allows us to focus our error analysis on output at ${\cal I}^+$.

Table \ref{tab:Conv_errorJ} gives the convergence rate of the error in $J$
measured at ${\cal I}^+$ at various times during
the $\ell=2$ run for the three methods
shown in Fig.~\ref{fig:InfErrJnorm}. Clean second order convergence, measured
either with an $L_2$ or $L_\infty$ norm, is indicated in all cases.
The corresponding convergence rates for the $\ell=3$ runs are given in
Table \ref{tab:Conv_errorJ_l3}.
The $\ell =2$ runs are more discriminating because $|J|$ has a $\sin^2
\theta$ dependence which peaks at the equator close to the interpatch
interpolation, as opposed to the $\sin^2 \theta \cos \theta$ dependence of the
$\ell=3$ case which vanishes at the equator.  In the following we restrict our
discussion to the $\ell=2$ case.

\begin{figure}[htp]
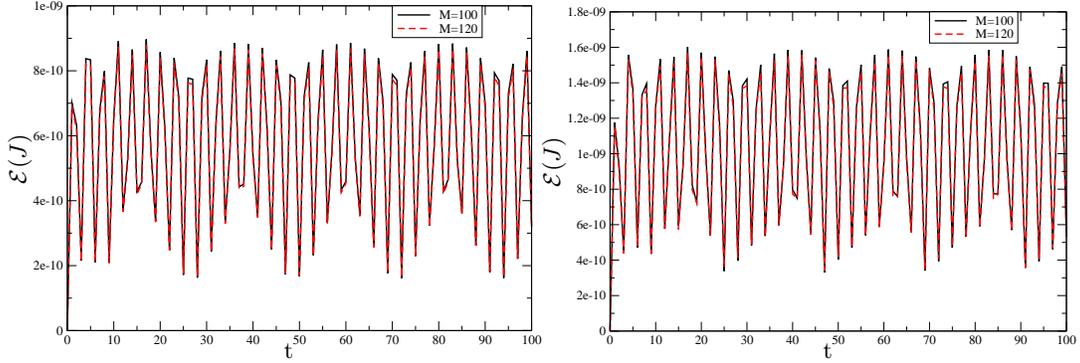
 %  figure placement: here, top, bottom, or page
    \centering
    \psfrag{xlabel}{t}
    \psfrag{ylabel}{${\cal E}(J)$}
    \includegraphics*[width=7cm]{Figure07} 
    \includegraphics*[width=7cm]{Figure08} 
    \caption{Plots of the error ${\cal E}(J)$ at ${\cal I}^+$ vs $t$, 
       for the circular patch with dissipation measured with the
      $L_2$ norm (left plot) and  the $L_\infty$ norm (right plot).
     The error for the $M=100$ grid is
     rescaled and overlaid on the error for the $M=120$ grid to exhibit
     the second order convergence. The smaller $L_2$ error
     indicates that the maximum error arises near the equator. }
    \label{fig:ErrJnorm}
 \end{figure}

\begin{figure}[htp] %  figure placement: here, top, bottom, or page
    \centering
    \includegraphics*[width=14cm]{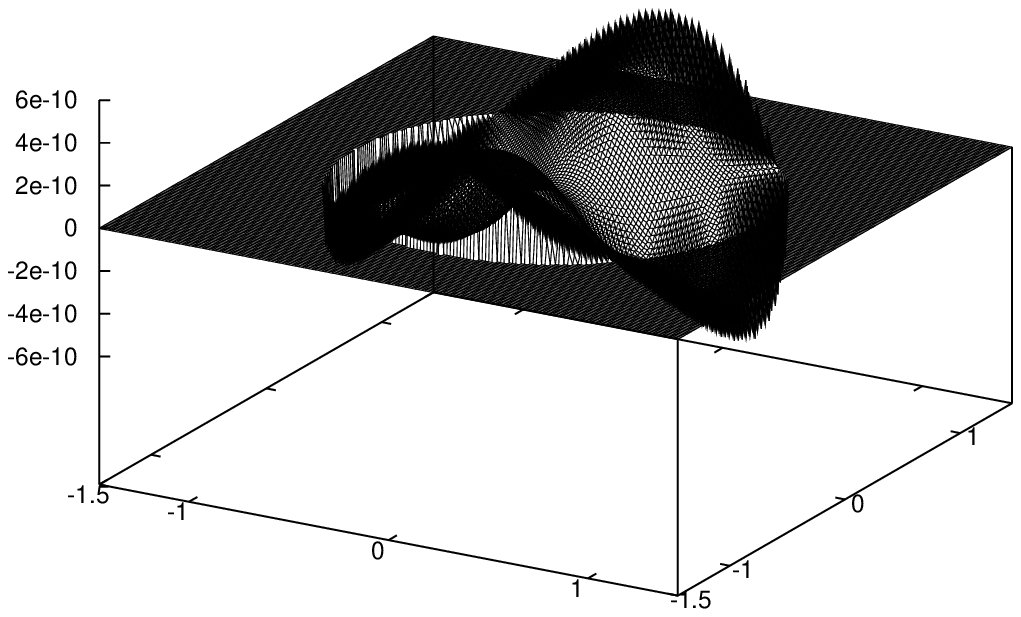} 
    \caption{Surface plot of the error in $J$ at $t=100$ for the
      circular patch run (with dissipation).}
    \label{fig:CrDissErrJSurf}
 \end{figure}

\begin{figure}[htp] %  figure placement: here, top, bottom, or page
    \centering
    \includegraphics*[width=14cm]{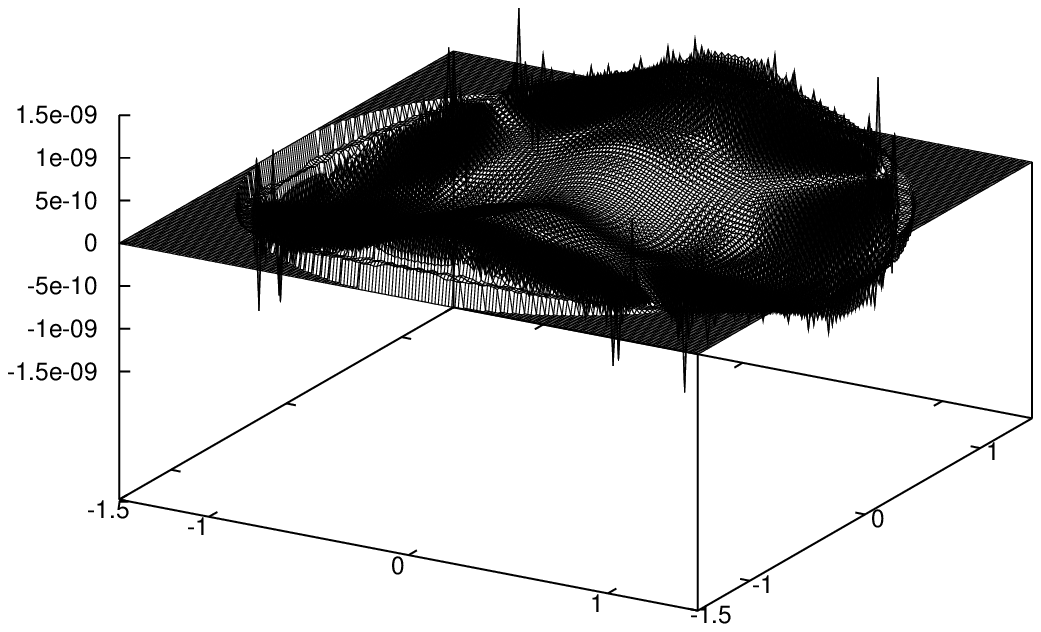} 
    \caption{Surface plot of the error in $J$ at $t=100$ for the circular
       patch run without dissipation.}
    \label{fig:CrErrJSurfEG}
 \end{figure}

\begin{figure}[htp] %  figure placement: here, top, bottom, or page
    \centering
    \includegraphics*[width=14cm]{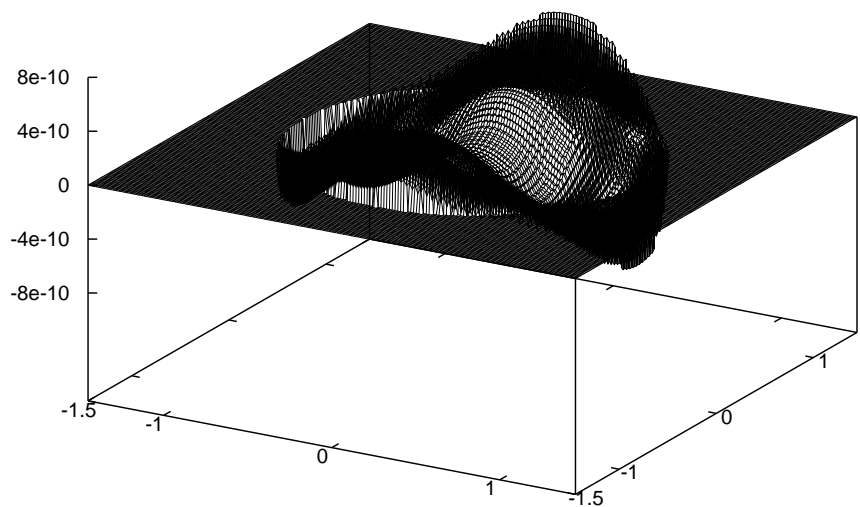} 
    \caption{Surface plot of the error in $J$ at $t=100$ for the square 
             patch run.}
    \label{fig:SqErrJSurf}
 \end{figure}

The time dependence of the $L_2$ and $L_\infty$ errors in $J$ at ${\cal I}^+$
for the circular patch run (with dissipation) is plotted in
Fig.~\ref{fig:ErrJnorm}. The plots are based upon output at integer values of
$t$, which samples the error at various phases during the underlying period
$T=2\pi$. The errors for the two grids used in the convergence measurements are
rescaled to the values for the finest $M=120$ grid, with the overlap again
confirming clean convergence. The magnitude of the error is approximately
$0.1\%$ the value of $J$. The $L_2$ error is
smaller than the $L_\infty$ because the error is sharply peaked near the equator.
This error
pattern in the North hemisphere is exhibited in the snapshot of
Fig.~\ref{fig:CrDissErrJSurf} at $t=100$. The profile is quite smooth - some of
the apparent jaggedness near the edge is an artificial effect of the irregular
pattern of grid points at the edge of the equator. The sharp spikes in the
corresponding error snapshot for the circular run without dissipation shown in
Figs.~\ref{fig:CrErrJSurfEG} illustrate the essential role of angular
dissipation in guarding the Northern hemisphere from the interpolation error at
the circular patch boundary. Such spikes are not apparent in the corresponding
error snapshot for the square patch shown in Fig.~\ref{fig:SqErrJSurf}. The more
regular square patch boundary does not require angular dissipation, although the
resulting error is larger than for the circular patch with dissipation.

\begin{table}[htp]
  \begin{center}
    \begin{tabular}[c]{|c|c|c|c|c|c|c|c|c|}
\hline
  Variable &
  circular patch &
  circular, no dissipation &
  square patch 

    \\ \hline \hline
%%%%%%
${\cal E}_{L_2}(J)_{t=1}$      & $2.01$ & $2.00$ & $2.01$ 
\\ \hline
%%%%%%
%%%%%%
${\cal E}_{L_2}(J)_{t=10}$      & $2.01$ & $1.98$ & $2.00$ 
\\ \hline
%%%%%%
%%%%%%
${\cal E}_{L_2}(J)_{t=90}$      & $2.00$ & $2.02$ & $2.02$ 
\\ \hline
%%%%%%
%%%%%%
${\cal E}_{L_2}(J)_{t=100}$      & $1.92$ & $2.03$ & $2.00$ 
\\ \hline
%%%%%%
%%%%%%
${\cal E}_{L_\infty}(J)_{t=1}$      & $2.01$ & $2.01$ & $2.01$ 
\\ \hline
%%%%%%
%%%%%%
${\cal E}_{L_\infty}(J)_{t=10}$      & $1.95$ & $2.00$ & $1.99$ 
\\ \hline
%%%%%%
%%%%%%
${\cal E}_{L_\infty}(J)_{t=90}$      & $2.07$ & $1.96$ & $2.00$ 
\\ \hline
%%%%%%
%%%%%%
${\cal E}_{L_\infty}(J)_{t=100}$      & $1.92$ & $2.01$ & $1.99$ 
\\ \hline
%%%%%%
\end{tabular}
    \caption{Convergence rates of the error ${\cal E}(J)$ at ${\cal I}^+$
             for the $\ell=2$ run,
             measured at times $t=1$, $t=10$, $t=90$, and $t=100$. 
          }
    \label{tab:Conv_errorJ}
  \end{center}
\end{table}

\begin{table}[htp]
  \begin{center}
    \begin{tabular}[c]{|c|c|c|c|c|c|c|c|c|}
\hline
  Variable &
  circular patch &
  circular, no dissipation &
  square patch 

    \\ \hline \hline
%%%%%%
${\cal E}_{L_2}(J)_{t=1}$      & $2.02$ & $2.01$ & $2.01$ 
\\ \hline
%%%%%%
%%%%%%
${\cal E}_{L_2}(J)_{t=10}$      & $2.00$ & $2.00$ & $2.00$ 
\\ \hline
%%%%%%
%%%%%%
${\cal E}_{L_2}(J)_{t=90}$      & $2.03$ & $2.02$ & $2.02$ 
\\ \hline
%%%%%%
%%%%%%
${\cal E}_{L_2}(J)_{t=100}$      & $2.05$ & $2.00$ & $2.01$ 
\\ \hline
%%%%%%
%%%%%%
${\cal E}_{L_\infty}(J)_{t=1}$      & $2.02$ & $2.02$ & $2.02$ 
\\ \hline
%%%%%%
%%%%%%
${\cal E}_{L_\infty}(J)_{t=10}$      & $1.99$ & $1.99$ & $2.00$ 
\\ \hline
%%%%%%
%%%%%%
${\cal E}_{L_\infty}(J)_{t=90}$      & $2.02$ & $2.02$ & $2.04$ 
\\ \hline
%%%%%%
%%%%%%
${\cal E}_{L_\infty}(J)_{t=100}$      & $2.00$ & $2.00$ & $1.99$ 
\\ \hline
%%%%%%
\end{tabular}
    \caption{Convergence rates of the error ${\cal E}(J)$ at ${\cal I}^+$
             for the $\ell =3$ run,
             measured at times $t=1$, $t=10$, $t=90$, and $t=100$. 
          }
    \label{tab:Conv_errorJ_l3}
  \end{center}
\end{table}

\clearpage

\subsection{Test results for the news}

We now compare test results for the news function in terms of a direct
calculation of $N$ via (\ref{eq:news}) and a calculation of $N_\Psi$ via
(\ref{eq:npsi}) using the Weyl component $\Psi$ given in (\ref{eq:psia}). We
restrict the discussion to the $\ell=2$ runs which are more challenging than
$\ell=3$ with respect to problems near the equator. Tables \ref{tab:Conv_errorN}
and \ref{tab:Conv_errorNewsPsi} give the convergence rates of the $L_2$ and
$L_\infty$ errors in $N$ and $N_\Psi$ measured at various times for runs with
the circular patch (with dissipation), the circular patch without dissipation
and the original square patch methods. At the final run time $t=100$,
measurements for all cases show clean second order convergence, although there
is a small departure in the $N_\Psi$ rates at early times.

\begin{figure}[htp]
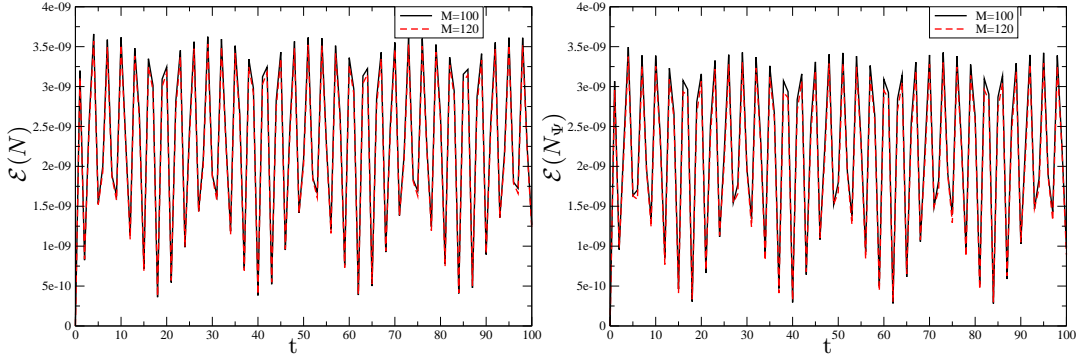
 %  figure placement: here, top, bottom, or page
    \centering
    \begin{psfrags}
    \psfrag{xlabel}{t}
    \psfrag{ylabel}{${\cal E}(N)$}
    \includegraphics*[width=7cm]{Figure12} 
    \end{psfrags}
    \begin{psfrags}
    \psfrag{xlabel}{t}
    \psfrag{ylabel}{${\cal E}(N_{\Psi})$}
    \includegraphics*[width=7cm]{Figure13} 
    \end{psfrags}
    \caption{Plots of the $L_2$ errors vs t for
             $N$ (left plot) and $N_\Psi$ (right plot) for the circular patch
              runs. The plots for the $M=100$ grid are rescaled to
	      the $M=120$ grid. The plots are based upon output at at integer
	      values of $t$.
             }
    \label{fig:ErrNewsnorm}
 \end{figure}

The plots of the $L_2$ error vs time  for the circular patch runs in
Fig.~\ref{fig:ErrNewsnorm} show little difference in the time behavior between
$N$ and $N_\Psi$, although the error in $N_\Psi$ is slightly smaller. The
$L_\infty$ errors measured at the end of the runs on the $M=120$ grid are given
in Table \ref{tab:InfNull_norminf} for the circular patch, the circular patch
without dissipation and the square patch. The best results are obtained for the
circular patch, which shows an $\approx 30\%$ improvement over the original
square patch treatment. The results also show the essential improvement due to
the use of angular dissipation. For the circular patch, the error in $N_\Psi$
was $\approx 24\%$ smaller than the error in $N$ at the end of the run. But it
is also clear from the plots of the $L_2$ error Fig.~\ref{fig:ErrNewsnorm} that
this ratio depends when and where this ratio is taken. At the equator where the
news takes its maximum value, its modulus for this test is $|N_{analytic}|
\approx 8\times 10^{-8}$. At the end of run, the corresponding fractional errors
in $N_\Psi$ and $N$ are $\approx 4\%$  for averaged values and $\approx 9\%$ for
the maximum errors at the equator.

\begin{figure}[htp] %  figure placement: here, top, bottom, or page
    \centering
    \includegraphics*[width=14cm]{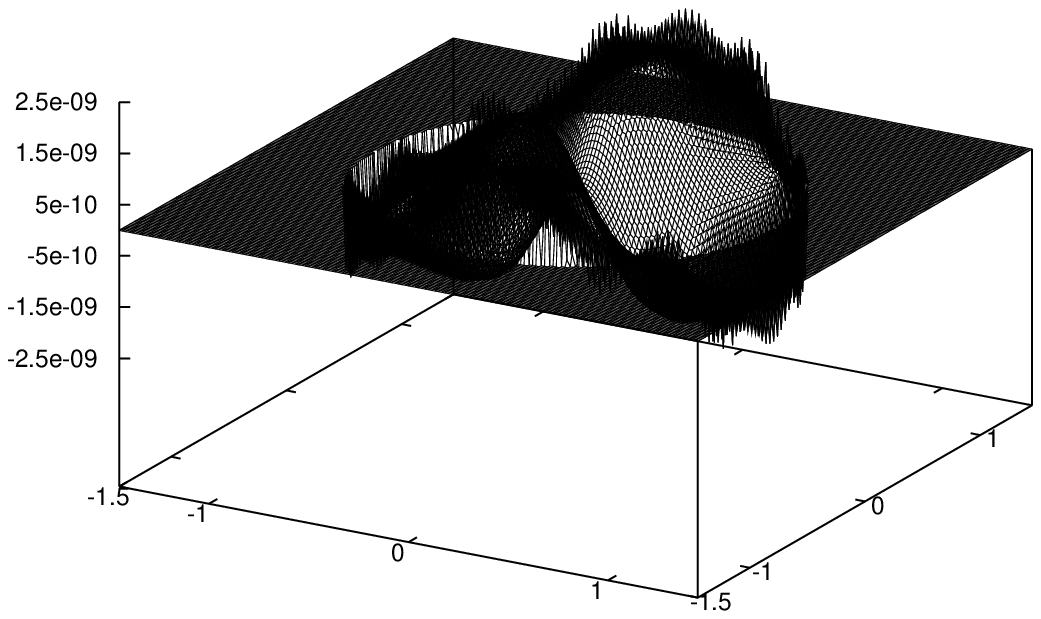} 
    \caption{Surface plot of the error in $N$ at $t=100$ for the 
             circular patch.}
    \label{fig:CrDissErrNewsSurf}
 \end{figure}

\begin{figure}[htp] %  figure placement: here, top, bottom, or page
    \centering
    \includegraphics*[width=14cm]{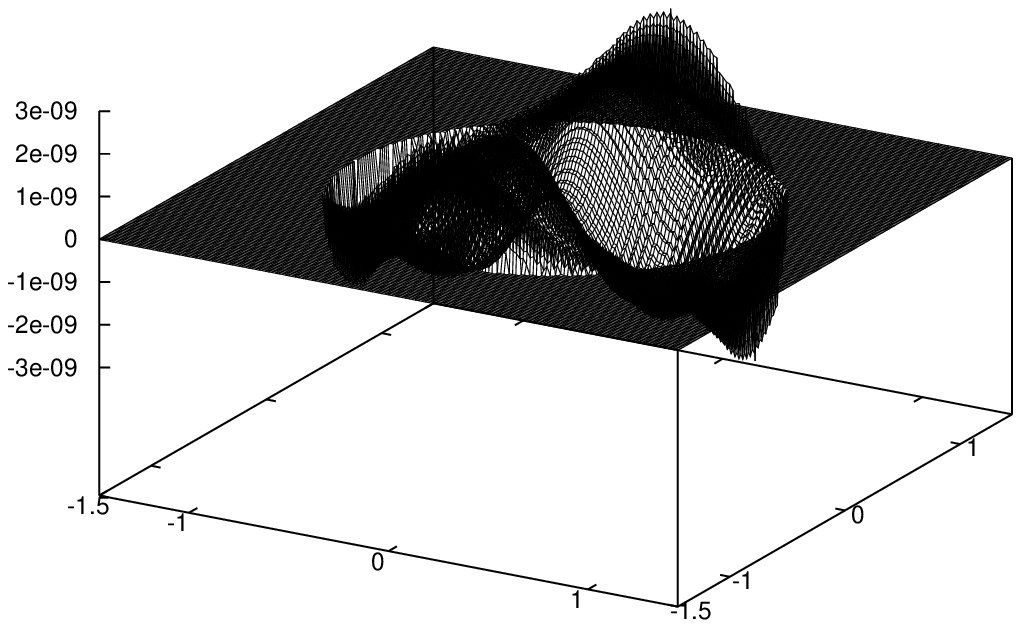} 
    \caption{Surface plot of the error in $N$ at $t=100$ for the
             square patch. }
    \label{fig:SqErrNewsSurf}
 \end{figure}

\begin{figure}[htp] %  figure placement: here, top, bottom, or page
    \centering
    \includegraphics*[width=14cm]{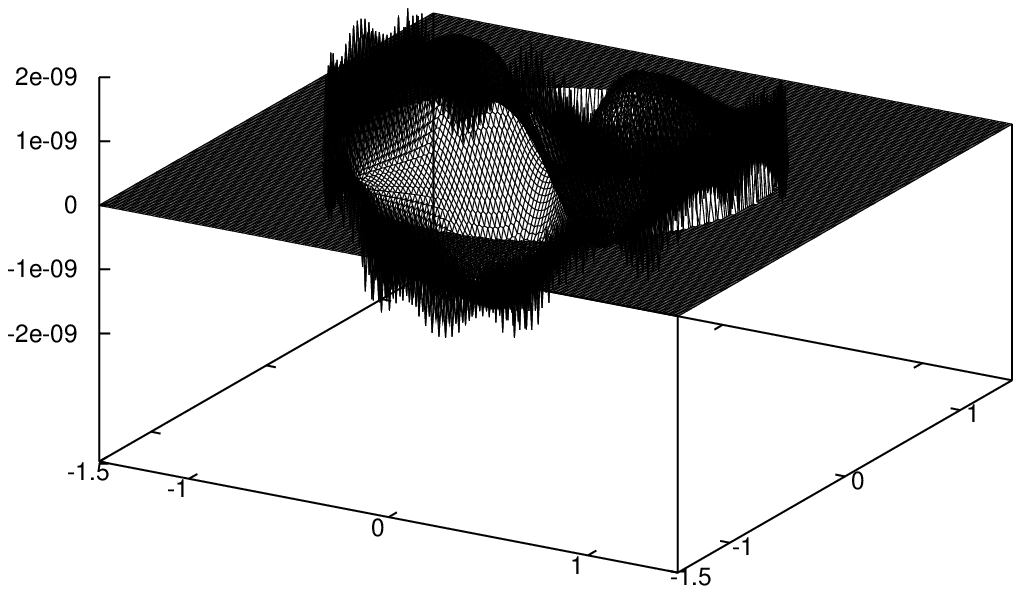} 
    \caption{Surface plot of the error in $N_{\Psi}$ at $t=100$ 
             for the circular patch.}
    \label{fig:CrDissErrNPsi4Surf}
 \end{figure}

\begin{figure}[htp] %  figure placement: here, top, bottom, or page
    \centering
    \includegraphics*[width=14cm]{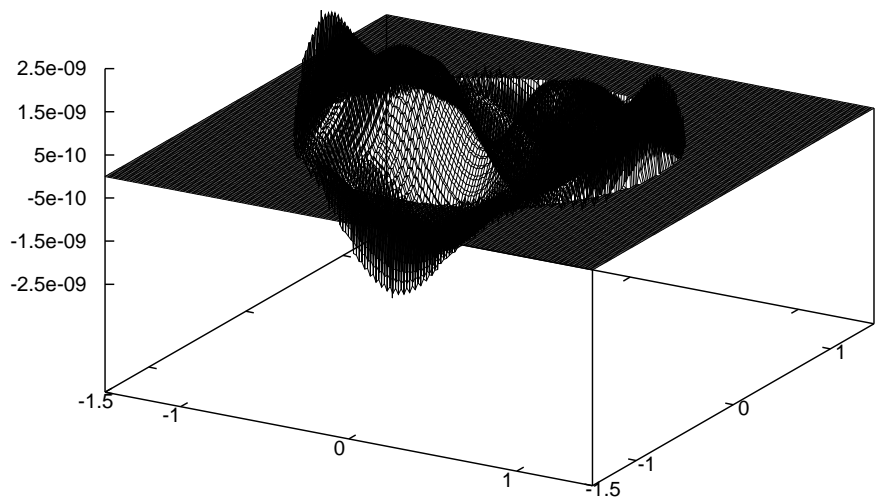} 
    \caption{Surface plot of the error in $N_{\Psi}$ at $t=100$ 
             for the square patch.}
    \label{fig:SqnoDErrNPsi4Surf}
 \end{figure}

Surface plots of the error in $N$ and $N_\Psi$ at the end of the run are given
in Figs.~\ref{fig:CrDissErrNewsSurf} - \ref{fig:SqnoDErrNPsi4Surf} for the
circular and square patches. The lack of sharp spikes in the errors for the
circular patches shows the effectiveness of applying angular dissipation. There
is slightly more jaggedness near the equator for the circular vs square patch
errors, but this is overbalanced by the relative smallness of the circular patch
error. 

\begin{table}[htp]
  \begin{center}
    \begin{tabular}[c]{|c|c|c|c|c|c|c|c|c|}
\hline
  Variable &
  circular patch &
  circular, no dissipation&
  square patch 

    \\ \hline \hline
%%%%%%
${\cal E}_{L_2}(N)_{t=1}$      & $2.05$ & $2.05$ & $2.05$ 
\\ \hline
%%%%%%
%%%%%%
${\cal E}_{L_2}(N)_{t=10}$      & $2.05$ & $2.05$ & $2.04$ 
\\ \hline
%%%%%%
%%%%%%
${\cal E}_{L_2}(N)_{t=90}$      & $2.04$ & $2.04$ & $2.01$ 
\\ \hline
%%%%%%
%%%%%%
${\cal E}_{L_2}(N)_{t=100}$      & $2.01$ & $2.07$ & $2.05$ 
\\ \hline
%%%%%%
%%%%%%
${\cal E}_{L_\infty}(N)_{t=1}$      & $2.04$ & $2.04$ & $2.04$ 
\\ \hline
%%%%%%
%%%%%%
${\cal E}_{L_\infty}(N)_{t=10}$      & $2.04$ & $1.99$ & $2.04$ 
\\ \hline
%%%%%%
%%%%%%
${\cal E}_{L_\infty}(N)_{t=90}$      & $2.01$ & $2.01$ & $2.06$ 
\\ \hline
%%%%%%
%%%%%%
${\cal E}_{L_\infty}(N)_{t=100}$      & $1.98$ & $2.00$ & $1.93$ 
\\ \hline
%%%%%%
\end{tabular}
    \caption{Convergence rates of the error ${\cal E}(N)$,
             measured at $t=1$, $t=10$, $t=90$, and $t=100$. 
          }
    \label{tab:Conv_errorN}
  \end{center}
\end{table}

\begin{table}[htp]
  \begin{center}
    \begin{tabular}[c]{|c|c|c|c|c|c|c|c|c|}
\hline
  Variable &
  circular patch &
  circular, no dissipation&
  square patch 

    \\ \hline \hline
%%%%%%
${\cal E}_{L_2}(N_{\Psi})_{t=1}$      & $2.11$ & $2.10$ & $2.11$ 
\\ \hline
%%%%%%
%%%%%%
${\cal E}_{L_2}(N_{\Psi})_{t=10}$      & $2.13$ & $2.13$ & $2.11$ 
\\ \hline
%%%%%%
%%%%%%
${\cal E}_{L_2}(N_{\Psi})_{t=90}$      & $2.09$ & $2.09$ & $2.08$ 
\\ \hline
%%%%%%
%%%%%%
${\cal E}_{L_2}(N_{\Psi})_{t=100}$      & $2.02$ & $1.98$ & $2.00$ 
\\ \hline
%%%%%%
%%%%%%
${\cal E}_{L_\infty}(N_{\Psi})_{t=1}$      & $2.08$ & $2.08$ & $2.08$ 
\\ \hline
%%%%%%
%%%%%%
${\cal E}_{L_\infty}(N_{\Psi})_{t=10}$      & $2.09$ & $2.05$ & $2.10$ 
\\ \hline
%%%%%%
%%%%%%
${\cal E}_{L_\infty}(N_{\Psi})_{t=90}$      & $2.05$ & $2.00$ & $2.06$ 
\\ \hline
%%%%%%
%%%%%%
${\cal E}_{L_\infty}(N_{\Psi})_{t=100}$      & $1.98$ & $2.01$ & $1.93$ 
\\ \hline
%%%%%%
\end{tabular}
    \caption{Convergence rates of the error ${\cal E}(N_{\Psi})$,
              measured at $t=1$, $t=10$, $t=90$, and $t=100$. 
          }
    \label{tab:Conv_errorNewsPsi}
  \end{center}
\end{table}

\begin{table}[htp]
  \begin{center}
    \begin{tabular}[c]{|c|c|c|c|}
\hline
  Variable &
  circular patch &
  circular, no dissipation&
  square patch 

    \\ \hline \hline
%%%%%
${\cal E}_{L_\infty}(N)$ & $2.247\times 10^{-9}$ & $3.325\times 10^{-9}$ & $2.897\times 10^{-9}$ 
\\ \hline
%%%%%
%%%%%
${\cal E}_{L_\infty}(N_{\Psi})$ & $1.706\times 10^{-9}$ & $2.747\times 10^{-9}$ & $2.315\times 10^{-9}$
\\ \hline
%%%%%
\end{tabular}
    \caption{The values of the $L_\infty$ errors in $N$ and $N_\Psi$
            measured at $t=100$,}
    \label{tab:InfNull_norminf}
  \end{center}
\end{table}

\clearpage

\section{Conclusion}
\label{sec:concl}

We have proposed two new methods for enhancing the accuracy of CCE, One is a
numerical method modifying the stereographic patches used in the
characteristic evolution code to conform to the circular patch boundaries as
used in meteorology~\cite{browning}.  The other is a geometrical method that
bases the waveform on the limiting behavior at ${\cal I}^+$ of the Weyl tensor
component $\Psi_4$ rather than the news function.

We have used a scalar wave testbed to compare the circular patch method against
the cubed sphere method, which is also extensively used in
meteorology~\cite{ronchi}. We found, for equivalent computational expense, that
the cubed-sphere method has an edge in accuracy over the stereographic method.
The cubed-sphere error in the scalar field ${\cal E}(\Phi)$ is $\approx
\frac{1}{3}$ the stereographic error but that the advantage is smaller for the
higher derivatives required in gravitational waveform extraction. The
cubed-sphere error ${\cal E}(\bar \eth \eth^2\Phi)$ is only $\approx
\frac{4}{5}$ the stereographic error. An advantage of the stereographic approach
is its relative programming simplicity. But as originally  pointed out
in~\cite{ronchi}, and demonstrated recently for the case of a characteristic
evolution code~\cite{roberto}, once all the necessary infrastructure for
interpatch communication is in place, the shared boundaries of the cubed-sphere
approach admit a highly scalable algorithm for parallel architectures. 

We used the circular patch stereographic code to compare waveform extraction
in a linearized wave test directly via the Bondi news function $N$
and its counterpart $N_\Psi$
constructed from the Weyl curvature. For this purpose, we were able to
successfully implement a new form of angular dissipation in the characteristic
evolution code, which otherwise would be prone to high frequency error
introduced by the irregular way a circular boundary cuts through a square grid.
Our test results show that this dissipation works: the
resulting error in the waveforms and metric quantities is smooth. In addition,
the extensive analytic and numerical manipulations carried out to compute the
limiting behavior of the Weyl curvature was demonstrated to yield second order
accurate results for $N_\Psi$.   
 
In the linearized tests presented here, neither $N$ nor $N_\Psi$ was a clear
winner. We already knew that the original news module based upon a square
stereographic patch worked well in the linear regime. The news $N$ calculated on
a circular patch had lower error than that on a square patch but only by a
$\approx 30\%$ factor. In turn, the news calculated via $N_\Psi$ on the circular
patch had a lower error than $N$ on the circular patch by a $\approx 24\%$
factor. Weyl tensor extraction is slightly more accurate than news function
extraction, even though there are many more terms involved.

All errors were second order convergent. However, while there was a small 
fractional error $\approx .1\%$ in metric quantities such as $J$, the
corresponding averaged error in the $N_\Psi$ and $N$ was $\approx 4\%$ 
for the circular patch runs and the maximum error at the equator was
$\approx 9\%$. These  errors did not vary appreciably ($\approx 30\%$)
with the choice of discretization method, i.e. circular patch or
square patch. They reflect the intrinsic difficulty in extracting waveforms due to
the delicate cancellation of leading order terms in the underlying metric and
connection when computing $O(1/r)$ quantities such as $\Psi_4$. The excellent
accuracy that we find for the metric suggests that perturbative waveform
extraction must suffer the same difficulty. In that case it is just less obvious
how to quantify the errors. The delicate issues involved at ${\cal I}^+$ have
been shown to have counterparts in extraction on a finite
worldtube~\cite{lehnmor}.

Waveforms are not easy to extract accurately. However, the convergence of our
error measurements is a positive sign that higher order finite difference
approximations might supply the accuracy that is needed for realistic
astrophysical applications. Whether the advantages the new methods proposed here
prove to be significant will depend upon the results of future application in
the nonlinear regime.

\centerline{\bf Acknowledgments} 

We thank Thomas Maedler for checking the calculations in Sec.~\ref{sec:weyl} and
G.~L. Browning for correspondence concerning the application of stereographic
patches in computational fluid dynamics. NTB thanks Max-Planck-Institut f\" ur
Gravitationsphysik, Albert-Einstein-Institut for hospitality; BS thanks
University of South Africa for hospitality; and MCB thanks University of
Pittsburgh for hospitality. We have benefited from the use of the Cactus
Computational Toolkit (http://www.cactuscode.org). Computer time was provided by
the Pittsburgh Supercomputing Center through a TeraGrid Wide Roaming Access 
Computational Resources Award, and we owe special thanks to R. G\'{o}mez for his
assistance. This work was supported by the Sherman Fairchild Foundation and the
National Science Foundation under grants PHY-061459 and PHY-0652995 to the
California Institute of Technology;  the National Science Foundation grant
PH-0553597 to the University of Pittsburgh; and by the National Research
Foundation, South Africa, under GUN 2075290.

\end{document}